\documentclass[letterpaper]{article}
\usepackage[preprint]{aaai2027}
\usepackage[hyphens]{url}
\usepackage{graphicx}
\usepackage{natbib}
\usepackage{caption}
\usepackage{amsmath}
\usepackage{array}
\usepackage{booktabs}
\usepackage{placeins}
\usepackage{tabularx}
\usepackage{xcolor}

\newcolumntype{Y}{>{\centering\arraybackslash}X}
\definecolor{trajectoryA}{HTML}{D9EAF4}
\definecolor{trajectoryB}{HTML}{F7E1C3}
\newcommand{\trajA}{\colorbox{trajectoryA}{\makebox[1.05em]{\strut A}}}
\newcommand{\trajB}{\colorbox{trajectoryB}{\makebox[1.05em]{\strut B}}}
\newenvironment{promptblock}[1]{%
  \par\smallskip\noindent\textbf{#1.}\par
  \begingroup\footnotesize
  \setlength{\parindent}{1em}\setlength{\parskip}{1pt}}
  {\par\endgroup}

\usepackage{etoolbox}
\makeatletter
\patchcmd{\@maketitle}
  {\footnote{Corresponding author\ifaaai@corrmulti{s}\fi.}}
  {\footnote{Corresponding author: b.heydari@northeastern.edu}}
  {}{\PackageWarning{preprint}{corresponding-author footnote patch FAILED}}
\makeatother

\title{Cheap Talk Stabilizes Strategic Interaction in LLM Agents}

\author{
    Nunzio Lor\`e,
    Hongan Zhu,
    Babak Heydari\corresponding
}
\affiliations{
    Multi-Agent Intelligent Complex Systems (MAGICS) Lab,
    Northeastern University, Boston, MA, USA
}

\begin{document}

\maketitle

\begin{abstract}
Large language models are increasingly deployed as interacting agents, making
the persistence of their action policies across repeated interaction critical
for reliable multi-agent operation. We investigate whether and how
agent-generated, non-binding pre-play communication (``cheap talk'') increases
such persistence in four open-weight 7--9B-parameter LLMs. Our
experiments span four repeated two-player games---Prisoner's Dilemma,
Snowdrift, Stag Hunt, and Harmony---with incentive structures ranging from
strategic conflict to alignment, each presented in six contexts. We observe
unstable trajectories in all four games, although their prevalence and
magnitude depend strongly on model and context. Across models, games, and
contexts, cheap talk is predominantly stabilizing, with five corrected
reversals concentrated in social or team framings; effects vary substantially
by model and context. Controlled current-message interventions identify two
separable output-level channels in Qwen: reduced action uncertainty and less
between-round drift in action probabilities. Matched history-by-message
counterfactuals further show that recent partner behavior conditions how
mutual-benefit versus self-prioritizing language affects policy persistence.
Finally, in Prisoner's Dilemma, we identify in Qwen and Falcon a
history-balanced policy-content direction in late transformer layers;
projecting out this direction increases realized switching during closed-loop
play, demonstrating that complete trajectories are causally sensitive to this
component. Together, these findings show that cheap talk can make individual
trajectories more persistent across diverse incentive structures, while
revealing that the magnitude and mechanisms of stabilization are model- and
history-dependent.
\end{abstract}

\section{Introduction}

Large language models (LLMs) increasingly act as autonomous or delegated agents in shared environments \cite{acharya2025agentic,shavit2023practices,sapkota2025ai}. This introduces failure modes less visible in single-turn evaluation \cite{li2024survey,han2024llm}, while multi-agent deployment can compound strategic misalignment with instability \cite{cemri2025multi}. Because one agent's action enters another's later context, small behavioral fluctuations can compound across rounds \cite{lore2024strategic}. Predictability is therefore important for reliable operation \cite{novikova2025consistency} and human expectations \cite{barak2025humans}.

We study \emph{behavioral stability} in the narrow sense of adjacent-round
action persistence. This trajectory-level property cannot be inferred from
population
averages: a stable aggregate share of one action can coexist with every
individual agent alternating between the two. Some switching is adaptive, since
an agent may reasonably respond to a change in its partner's behavior. The
oscillation we observe is largely not of that kind: it reaches rates that no
memoryless randomization over actions can produce, and it persists even under incentive
structures in which one action strictly dominates the other. We therefore use
adjacent-round switching as an operational measure of first-order trajectory
persistence, not as a claim that every action change is irrational or that the
metric captures every form of predictability.

Pre-play communication is a natural intervention. \emph{Cheap talk} is costless and non-binding, so it can change what agents express about their intentions without changing incentives \cite{crawford1982strategic,farrell1996cheap}. Neither cheap-talk theory nor the human experimental record fixes what to expect from LLM agents: their messages may convey strategic intent, but text may equally sharpen action distributions, anchor policies, or be read against recent play. We therefore ask whether cheap talk stabilizes repeated interaction across incentive structures, and through which mechanisms.

\begin{figure*}[t]
\centering
\includegraphics[width=\textwidth]{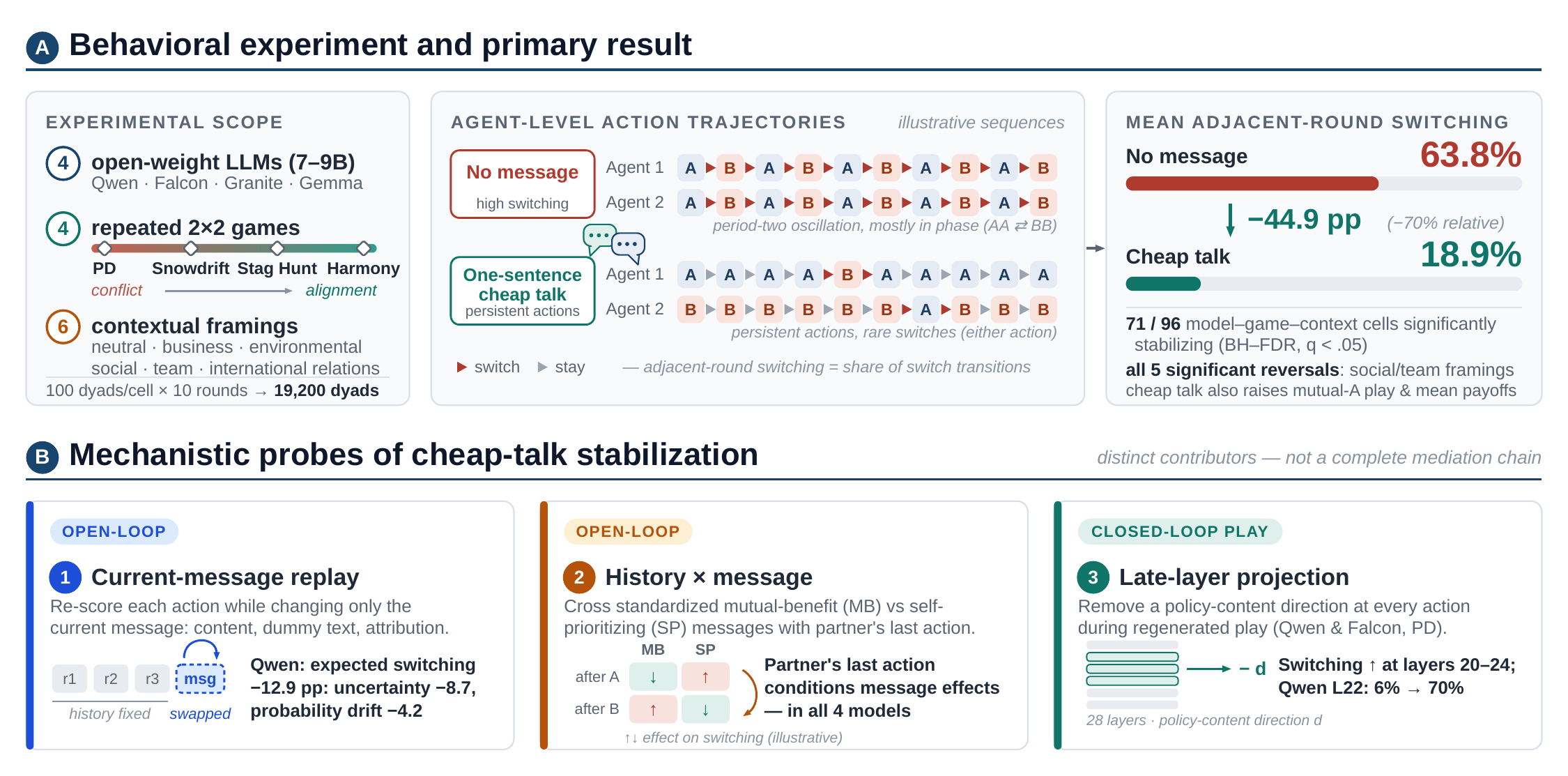}
\caption{Design, primary result, and mechanistic follow-ups. (A) Four open-weight 7--9B LLMs play four repeated $2\times2$ games under six framings, with or without one-sentence pre-play cheap talk; illustrative trajectories contrast period-two cycling with persistent play. Across the 96 model--game--context comparisons, mean adjacent-round switching falls from 63.8\% to 18.9\%. (B) Three analyses examine current-message effects along fixed histories, history-conditioned message effects, and the sensitivity of regenerated trajectories to projecting out a late-layer policy-content direction.}
\label{fig:overview}
\end{figure*}

We evaluate four open-weight 7--9B models in repeated Prisoner's Dilemma, Snowdrift, Stag Hunt, and Harmony. Across six framings and two treatments, 100 independent ten-round dyads per model--game--context--treatment cell yield 19,200 repeated-game dyads. Figure~\ref{fig:overview} summarizes the design and the follow-ups.

Cheap talk stabilizes 71 of the 96 model--game--context comparisons after correction, while five reverse in social or team framings. Three follow-ups then separate fixed-history policy effects (Experiments 1--2) from closed-loop sensitivity to a late-layer component (Experiment 3). The contribution is therefore not that communication shifts average behavior, but that it makes individual trajectories more persistent, and that this effect separates into output-level and representation-level components.

\section{Related Work}

\subsection{Strategic Decision-Making in LLMs}

Strategic competence requires representing incentives, updating beliefs, and comparing contingent actions in an interactive environment, and remains challenging for LLMs \cite{zhu2025gtalign}. Early work proposed LLMs as scalable ``in silico'' proxies for human participants \cite{horton2023large}, but behavior in adversarial and game-theoretic settings is often boundedly rational and heuristic \cite{sun2025game}, diverging from both rational-choice predictions and established behavioral regularities \cite{kitadai2023toward,zhang2024llm,guo2023gpt,mei2024turing,fontana2025nicer}. Models also struggle to refine beliefs from interaction evidence and apply humanlike strategies rigidly \cite{fan2024can,zheng2025beyond}, so the validity of LLM-based strategic simulation depends on experimental design and opaque training choices as much as on capability \cite{wang2025limits,willis2025will}. A parallel literature seeks to steer multi-agent behavior toward cooperation or prosocial outcomes \cite{justus2025llms,chen2024instigating,tran2025multi,wang2025beyond,piatti2024cooperate}, emphasizing welfare or equilibrium rather than whether an individual policy persists across rounds. This distinction matters because a change in aggregate cooperation can reflect either a genuine policy shift or greater stochastic dispersion around an unchanged tendency.

\subsection{Behavioral Stability}

LLM instability reflects probabilistic sampling and computational nondeterminism \cite{atil2024non}, pre-training and system design \cite{wang2025assessing,roy2025interpreting}, and sensitivity to prompt wording, framing, and contextual cues \cite{huang2024far,mozikov2024eai,errica2025did,sclar2023quantifying}. Greater scale does not necessarily eliminate the problem \cite{zhou2024larger,yu2025performance}, creating concern where reproducible reasoning and dependable decisions are required \cite{carandang2025llms,blair2025llms,purushothama2025not,gallagher2024assessing}, and recent benchmarks accordingly evaluate reliability across repeated runs, perturbations, and scoring dimensions \cite{anghel2025diagnosing,nalbandyan2025score,jang2025rcscore}. Multi-turn interaction adds a feedback problem, since earlier outputs enter later contexts and consistency may deteriorate over an exchange \cite{li2025firm}, and context dependence is pronounced in behavioral and strategic tasks \cite{kovavc2024stick}. Our focus complements cross-run reliability by measuring within-trajectory policy persistence, which stable population averages can conceal and which is separate from whether the persistent action is cooperative. It also differs from consistency across semantically equivalent prompts: the unit of analysis is one agent's evolving policy within one repeated interaction.

\subsection{Strategic Communication}

Strategic communication has been studied through signaling, disclosure, persuasion, and related frameworks \cite{spence1978job,akerlof1978market,grossman1981informational,milgrom1981good,schelling1980strategy,kamenica2011bayesian}. Most relevant here is cheap talk: costless, non-binding communication without direct payoff effects or enforceable truthfulness, which admits uninformative babbling equilibria but can be informative when interests are sufficiently aligned \cite{crawford1982strategic,farrell1996cheap}. Credibility may also emerge from reliable behavioral histories, extended conversation, reputation, and repeated interaction \cite{sobel1985credibility,aumann2003long,golosov2014dynamic,blume2020strategic,wilson1997liar,lima2024infinitely,arechar2017m}. Human behavior departs systematically from these benchmarks: senders overcommunicate \cite{cai2006overcommunication,sanchez2007experimental,lafky2022preferences,wang2010pinocchio}, while receivers may overweight biased messages or discount information too strongly \cite{li2022we,kartik2007credulity,li2016cheap}, and whether LLMs reproduce, amplify, or attenuate these biases remains under study \cite{tjuatja2024llms,geva2025llms}. Less is known about whether cheap talk makes repeated policies more persistent across conflict and coordination environments, and whether any stabilization reflects strategic content, generic text, attribution, or interaction with observed behavior. Our design treats these as separable hypotheses rather than interpreting the presence of a message as a unitary intervention.

\subsection{Causal Analysis of Internal Representations}

Diagnostic probes show that information is decodable without establishing that a model uses it to produce an output \cite{hewitt2019designing,belinkov2022probing,elazar2021amnesic}. Causal mediation, tracing, patching, and steering intervene on internal states and measure behavioral change \cite{vig2020investigating,meng2022locating,zhang2024activation,li2023inference,rimsky2024steering}, while projection-based removal provides a complementary necessity test and causal abstraction a broader interpretive framework \cite{elazar2021amnesic,dobrzeniecka2025improving,geiger2025causal}. In multi-agent play, however, an altered action changes the history conditioning later decisions, so an intervention that changes one logit without regenerating subsequent play may miss the system-level consequence. We therefore combine matched output counterfactuals with closed-loop projection ablation, so that trajectory-level claims concern behavioral use rather than decodability alone.

\section{Methods}

\subsection{Experimental Design}

In each repeated symmetric $2\times2$ game, both agents choose A or B. Following standard notation, $R$ is each agent's payoff under AA, $P$ is each payoff under BB, $T$ is the payoff to B against A, and $S$ is the payoff to A against B. The games differ only in rank order: Prisoner's Dilemma ($T>R>P>S$), Snowdrift ($T>R>S>P$), Stag Hunt ($R>T>P>S$), and Harmony ($R>S>T>P$). Values 5, 3, 1, and 0 are assigned in rank order, holding scale fixed while varying incentives. Prompts show the payoff matrix but not the conventional game name. Each game is framed as neutral, business, environmental, social, team, or international-relations interaction.

For each model--game--context--treatment cell, we generate 100 independent ten-round dyads; agents are not told the horizon. Message calls execute sequentially for implementation, but the protocol is simultaneous in information: each agent composes from its history through the preceding round, the first message is buffered, and both messages are revealed before either action. Messages are non-binding and carry no truthfulness requirement, distinguishing cheap talk from commitment. The no-communication treatment supplies only the game and observed history. System prompts contain the framing, action definitions, and payoff matrix; round prompts report prior actions and payoffs, elicit and buffer messages when applicable, reveal the counterpart's message, and request A or B. Communication therefore changes pre-action text without changing payoffs or introducing within-round sequential persuasion.

\subsection{Behavioral Stability Metric}

Let $a_{idt}\in\{A,B\}$ be agent $i$'s action in dyad $d$ at round $t$. Trajectory instability is
\begin{equation}
S_{id}=\frac{1}{T-1}\sum_{t=2}^{T}\mathbf{1}\!\left\{a_{idt}\neq a_{id,t-1}\right\},\qquad T=10.
\end{equation}
Each complete dyad contributes 18 transitions, nine from each agent, and condition-level rates pool these indicators. The communication effect is
\begin{equation}
\Delta S=\bar S_{\mathrm{no\ message}}-\bar S_{\mathrm{message}},
\end{equation}
so positive values indicate stabilization. The metric is label-symmetric and agnostic about which action persists; stable A and stable B trajectories receive the same score.

Independent draws from a memoryless Bernoulli policy have expected switching
$2p(1-p)\leq 1/2$; near-ceiling rates therefore imply systematic temporal
structure rather than independent mixing. More generally, for a stationary binary series with marginal $p$ and first-order autocorrelation
$\rho_1$,
\begin{equation}
S=2p(1-p)\,(1-\rho_1),
\label{eq:autocorr}
\end{equation}
so at fixed $p$ the switch rate is an affine rescaling of $\rho_1$. Because
$2p(1-p)\leq 1/2$ for every $p$, any $S>1/2$ implies $\rho_1<0$ irrespective of
the marginal, whereas below one half the sign of $\rho_1$ depends on $p$; we
therefore draw this inference in one direction only. We report $S$ rather than
mean run length because run length is censored at the ten-round horizon,
whereas a per-transition rate is not.

A deterministic period-two sequence can nevertheless have $S=1$ while remaining
perfectly predictable from round parity, and period-two cycling is in fact the
second-order structure we observe. Under our rank-order payoffs, constant A
weakly dominates sustained in-phase and antiphase alternation in every game and
strictly dominates except antiphase Snowdrift, which ties. We therefore
interpret $S$ as first-order action persistence, characterizing the sign and
strength of adjacent-round dependence rather than general predictability,
strategic quality, or welfare. With nine transitions per agent, it estimates
that dependence within a short window and does not distinguish a settled cycle
from a transient that has not yet converged.

\subsection{Model Selection}

We evaluate four general-purpose open-weight models in a narrow 7--9B parameter range: Qwen 2.5 7B, Falcon 3 7B, Granite 3.3 8B, and Gemma 2 9B. Holding scale approximately constant limits parameter-count variation while retaining differences in training history and architecture, and open weights expose token-level action probabilities and internal activations. Moderate size makes 19,200 dyads, matched replays, and closed-loop interventions tractable; we exclude reasoning-distilled checkpoints to preserve a common class of instruction-tuned models. Compact models are also practically relevant because repeated multi-agent inference magnifies token and compute costs \cite{Zhang2025MaAS,Belcak2025Small}.

\subsection{Model Execution and Statistical Analysis}

Primary simulations use \texttt{ollama} at temperature 0.8, held fixed across treatments. This setting exposes variation in the sampled policy rather than evaluating only modal actions and is comparable to prior behavioral and multi-agent studies \cite{aher2023using,lore2024strategic}. Exact model artifacts, prompts, decoding and parsing rules, replay construction, and provenance are documented in Supplementary Material, Sections~A.3--A.5 and F.

The complete-case analysis excludes 39 dyads (0.20\%) with generation or parsing errors, including 25 partial dyads, leaving 19,161 dyads and 344,898 adjacent-round transitions. A sensitivity analysis retaining fallback actions and all available transitions from partial trajectories changes none of the 96 corrected significance classifications.

For each cell, a nonparametric bootstrap resamples whole dyads separately by treatment, preserving dependence between agents and rounds, and recomputes $\Delta S$ 10,000 times \cite{efron1979bootstrap}; we report percentile 95\% intervals and two-sided bootstrap $p$-values. Benjamini--Hochberg correction is applied across the 96 cell-level comparisons, with $q<.05$ defining a corrected result \cite{benjamini1995controlling}. Model-, game-, and overall estimates weight their constituent cells equally and use the same stratified bootstrap.

\subsection{Mechanistic Follow-Up}

Experiments 1 and 2 replay 20 communication-treatment dyads per model--game--context cell under fixed realized histories; Experiment 3 uses a separate Qwen/Falcon cohort and regenerates trajectories under activation intervention. Figure~\ref{fig:overview}B summarizes these complementary designs: the first two estimate open-loop policy effects, whereas the third captures closed-loop propagation.

\subsubsection{Experiment 1: Current-Message Policy Replay}

We rescore each action prompt under seven matched conditions: endogenous partner message; focal-agent text attributed to the partner; partner text attributed to the focal agent; focal agent's public message; unrelated dummy text; partner-formatted dummy text; and no message. Only the current block and its attribution change.

Let $p_t=P(A)$ under a counterfactual condition. Expected disagreement between
independent adjacent policy draws is
\begin{align}
E_t
&=p_t(1-p_{t+1})+(1-p_t)p_{t+1} \nonumber\\
&=p_t+p_{t+1}-2p_tp_{t+1} \nonumber\\
&=\underbrace{(p_{t+1}-p_t)^2}_{\text{probability drift}}
+\underbrace{p_t(1-p_t)+p_{t+1}(1-p_{t+1})}_{\text{action uncertainty}}.
\label{eq:expected-switch-decomposition}
\end{align}
This exact, label-symmetric decomposition separates movement in $P(A)$ from the adjacent policies' Bernoulli variances. It summarizes the open-loop policy along fixed histories and is distinct from realized switching in regenerated trajectories. Real versus dummy content under partner framing tests content; the same dummy text with and without partner formatting tests attribution; reattribution tests apparent source.

\subsubsection{Experiment 2: History-Conditioned Message Effects}

We cross the partner's most recent action (A or B) with a standardized mutual-benefit (MB) or self-prioritizing (SP) message. The outcome remains open-loop expected switching. The estimand is one-half of the history-by-message interaction: the SP-minus-MB contrast after A minus the same contrast after B. Halving places it on the scale of one conditional contrast, comparable to Experiment 1. A nonzero value means that recent partner behavior conditions the stability effect of message framing; the design does not assign a game-invariant semantic consistency label to any A/B--message pairing.

\subsubsection{Experiment 3: Layerwise Causal Intervention}

We intervene in Prisoner's Dilemma for Qwen and Falcon, the only depth-matched pair: both have 28 transformer layers, versus 40 for Granite and 42 for Gemma. Matching depth allows comparison at identical absolute layer indices and depth fractions without introducing total depth as a further difference. We test layers $2,4,\ldots,28$, use ten trajectories per context, and split complete decisions 65/35 into construction and held-out sets. At each layer, final-token residual means for mutual-benefit and self-prioritizing messages are balanced over partner-A and partner-B histories; their normalized difference defines a history-balanced \emph{policy-content} direction $d_l$ \cite{li2023inference,rimsky2024steering}. Balancing prevents unequal composition of the partner's latest action from driving the contrast, and the held-out split prevents the same decision from constructing and evaluating the direction.

We first add $d_l$, or an equal-norm random direction, on 150 held-out no-message replay prompts balanced across the six contexts and measure the A--B logit-margin slope. We then project out the candidate component during every action choice,
\begin{equation}
h'_{l,-1}=h_{l,-1}-(h_{l,-1}^{\top}d_l)d_l,
\end{equation}
while leaving message generation unchanged. Each ablated trajectory is paired with a clean trajectory using the same context and seed. The paired change in switching tests closed-loop sensitivity rather than mere decodability \cite{elazar2021amnesic,geiger2025causal,dobrzeniecka2025improving}.

\section{Results}

\begin{table*}[t]
\centering
\scriptsize
\setlength{\tabcolsep}{2.2pt}
{\renewcommand{\arraystretch}{0.88}
\begin{tabular*}{\textwidth}{@{\extracolsep{\fill}}llrrrrcccccc@{}}
\toprule
& & \multicolumn{2}{c}{\textbf{Switch rate (\%)}} & \multicolumn{2}{c}{\textbf{Pooled effect}} & \multicolumn{6}{c}{\textbf{FDR by context}} \\
\cmidrule(lr){3-4}\cmidrule(lr){5-6}\cmidrule(lr){7-12}
\textbf{Model} & \textbf{Game} & \textbf{No msg.} & \textbf{Msg.} & $\boldsymbol{\Delta S}$ \textbf{(pp)} & \textbf{95\% CI} & \textbf{N} & \textbf{B} & \textbf{E} & \textbf{S} & \textbf{T} & \textbf{IR} \\
\midrule
\textbf{Qwen 2.5 7B} & PD        & 85.2 & 13.1 & \textbf{72.1}\textsuperscript{***} & $\lbrack 69.5, 74.8\rbrack$ & $\mathbf{+}$ & $\mathbf{+}$ & $\mathbf{+}$ & $\mathbf{+}$ & $\mathbf{+}$ & $\mathbf{+}$ \\
             & Snowdrift & 99.5 & 12.6 & \textbf{86.9}\textsuperscript{***} & $\lbrack 84.9, 88.8\rbrack$ & $\mathbf{+}$ & $\mathbf{+}$ & $\mathbf{+}$ & $\mathbf{+}$ & $\mathbf{+}$ & $\mathbf{+}$ \\
             & Stag Hunt & 93.3 & 10.1 & \textbf{83.2}\textsuperscript{***} & $\lbrack 81.1, 85.3\rbrack$ & $\mathbf{+}$ & $\mathbf{+}$ & $\mathbf{+}$ & $\mathbf{+}$ & $\mathbf{+}$ & $\mathbf{+}$ \\
             & Harmony   & 99.8 &  9.3 & \textbf{90.5}\textsuperscript{***} & $\lbrack 88.7, 92.2\rbrack$ & $\mathbf{+}$ & $\mathbf{+}$ & $\mathbf{+}$ & $\mathbf{+}$ & $\mathbf{+}$ & $\mathbf{+}$ \\
\midrule
\textbf{Granite 3.3 8B} & PD        & 95.2 & 23.6 & \textbf{71.6}\textsuperscript{***} & $\lbrack 69.1, 74.1\rbrack$ & $\mathbf{+}$ & $\mathbf{+}$ & $\mathbf{+}$ & $\mathbf{+}$ & $\mathbf{+}$ & $\mathbf{+}$ \\
               & Snowdrift & 98.5 & 19.1 & \textbf{79.3}\textsuperscript{***} & $\lbrack 77.1, 81.5\rbrack$ & $\mathbf{+}$ & $\mathbf{+}$ & $\mathbf{+}$ & $\mathbf{+}$ & $\mathbf{+}$ & $\mathbf{+}$ \\
               & Stag Hunt & 80.5 & 14.7 & \textbf{65.8}\textsuperscript{***} & $\lbrack 62.3, 69.1\rbrack$ & $\mathbf{+}$ & $\mathbf{+}$ & $\mathbf{+}$ & $\mathbf{+}$ & $\mathbf{+}$ & $\mathbf{+}$ \\
               & Harmony   & 92.7 & 15.5 & \textbf{77.2}\textsuperscript{***} & $\lbrack 74.5, 79.9\rbrack$ & $\mathbf{+}$ & $\mathbf{+}$ & $\mathbf{+}$ & $\mathbf{+}$ & $\mathbf{+}$ & $\mathbf{+}$ \\
\midrule
\textbf{Gemma 2 9B} & PD        & 44.0 & 21.5 & \textbf{22.6}\textsuperscript{***} & $\lbrack 18.4, 26.6\rbrack$ & $\mathbf{+}$ & $\mathbf{+}$ & $\mathbf{+}$ & $\cdot$ & $\cdot$ & $\mathbf{+}$ \\
            & Snowdrift & 53.8 & 25.8 & \textbf{28.0}\textsuperscript{***} & $\lbrack 24.0, 32.1\rbrack$ & $\mathbf{+}$ & $\mathbf{+}$ & $\mathbf{+}$ & $\cdot$ & $\cdot$ & $\mathbf{+}$ \\
            & Stag Hunt & 16.7 & 11.5 & \textbf{5.2}\textsuperscript{**} & $\lbrack 1.9, 8.6\rbrack$ & $\mathbf{+}$ & $\mathbf{+}$ & $\cdot$ & $\mathbf{-}$ & $\mathbf{-}$ & $\cdot$ \\
            & Harmony   & 13.4 & 11.0 & 2.4 & $\lbrack -0.7, 5.6\rbrack$ & $\mathbf{+}$ & $\mathbf{+}$ & $\cdot$ & $\mathbf{-}$ & $\cdot$ & $\mathbf{+}$ \\
\midrule
\textbf{Falcon 3 7B} & PD        & 52.3 & 35.5 & \textbf{16.8}\textsuperscript{***} & $\lbrack 12.8, 20.8\rbrack$ & $\mathbf{+}$ & $\mathbf{+}$ & $\mathbf{+}$ & $\mathbf{+}$ & $\cdot$ & $\mathbf{+}$ \\
             & Snowdrift & 42.5 & 30.3 & \textbf{12.2}\textsuperscript{***} & $\lbrack 8.2, 16.1\rbrack$ & $\mathbf{+}$ & $\cdot$ & $\mathbf{+}$ & $\mathbf{-}$ & $\cdot$ & $\mathbf{+}$ \\
             & Stag Hunt & 25.1 & 24.6 & 0.5 & $\lbrack -3.1, 4.1\rbrack$ & $\cdot$ & $\cdot$ & $\cdot$ & $\cdot$ & $\cdot$ & $\cdot$ \\
             & Harmony   & 27.9 & 24.1 & \textbf{3.8}\textsuperscript{*} & $\lbrack 0.2, 7.4\rbrack$ & $\mathbf{+}$ & $\cdot$ & $\mathbf{+}$ & $\mathbf{-}$ & $\cdot$ & $\cdot$ \\
\bottomrule
\end{tabular*}}
\caption{Communication effects on adjacent-round switching. Model--game rows pool six contexts; positive $\Delta S$ indicates stabilization. Bold pooled effects are significant (\textsuperscript{*}$p<.05$, \textsuperscript{**}$p<.01$, \textsuperscript{***}$p<.001$). Context symbols show Benjamini--Hochberg results across 96 cells ($q<.05$): $\mathbf{+}$ stabilizing, $\mathbf{-}$ destabilizing, $\cdot$ null. Codes: neutral (N), business (B), environmental (E), social (S), team (T), and international relations (IR). Confidence intervals use 10,000 dyad-level bootstrap resamples.}
\label{standard}
\end{table*}

\subsection{Communication Reduces Action Switching}

Table~\ref{standard} summarizes the primary result. Equally weighting all 96
cells, switching is 63.8\% without communication and 18.9\% with it, a
reduction of 44.9 percentage points (95\% CI [44.1, 45.6]) or 70.4\% relative.
The direction is stabilizing in 81 cells, and 71 remain significant after
correction across the full family of tests.

The effect spans all games: pooled reductions are 45.8 points in Prisoner's Dilemma, 51.6 in Snowdrift, 38.7 in Stag Hunt, and 43.5 in Harmony, with every game-level interval excluding zero. After correction across all 96 comparisons, significantly stabilizing cells number 21/24, 19/24, 14/24, and 17/24, respectively. Falcon in Stag Hunt is the weakest model--game combination ($0.5$ points; interval includes zero).

Because action persistence is distinct from strategic quality, we also examine
what becomes persistent. Across games, communication raises Action-A rates by
18.49--23.49 percentage points, AA outcomes by 25.90--30.08 points, and mean
payoff by 0.404--1.036 points per agent per round. Stabilization is therefore
accompanied by a systematic shift toward persistent AA play and higher realized
scores rather than arbitrary persistence on either action (Supplementary
Material, Section~A.1).

\subsection{Model and Context Heterogeneity}

Qwen and Granite show the broadest effects: all 24 cells stabilize, with model-level reductions of 83.2 and 73.5 percentage points. Their no-communication switching rates are also highest (94.4\% and 91.7\%), strongly shaping the grand mean. Gemma and Falcon decline by 14.5 and 8.3 points; across their 48 cells, 23 stabilize, five destabilize, and 20 are null after correction, although both aggregate intervals exclude zero.

The near-ceiling baselines reflect population-wide period-two alternation. Of 4,800 Qwen no-communication agent trajectories, 91.5\% switch on all nine transitions, as do 83.2\% of Granite's 4,798; 92.3\% and 89.8\% switch on at least eight. In Harmony, both agents alternate in 591/600 Qwen and 498/600 Granite dyads; 73.4\% and 59.0\% of these are in phase, cycling jointly between AA and BB rather than taking turns in asymmetric outcomes. In Snowdrift, both agents alternate perfectly in 583/600 Qwen and 526/599 complete Granite dyads; 88.2\% and 79.7\% of these are likewise in phase. Thus, the Snowdrift exception in which antiphase alternation ties constant A at $(T+S)/2=R$ applies to only 11.8\% and 20.3\% of perfectly alternating dyads. The predominance of in-phase play also excludes mutual reciprocity: tit-for-tat against a tit-for-tat partner is absorbing from a symmetric opening, giving $S=0$, and yields antiphase alternation from an asymmetric one, so it cannot produce in-phase joint cycling from any start.

These trajectories are also anti-persistent in a sense that requires no knowledge of the action marginals: an agent switching on at least eight of nine transitions has $S\geq 8/9$, and since $2p(1-p)\leq 1/2$ for every $p$, Equation~\ref{eq:autocorr} gives $\rho_1<0$ without any estimate of its action frequencies.
Communication moves every model--game cell below the one-half threshold (Table~\ref{standard}), removing this signature. A simple preference for one action label also cannot explain the pattern, because it would predict persistence on the favored label rather than symmetric alternation. A tendency not to reproduce the most recently emitted action would be consistent with what we observe, since it is label-symmetric and predicts period-two play; separating it from strategic instability requires varying how prior rounds are presented, which we do not do here. The full audit appears in Supplementary Material, Section~A.2.

Communication is not universally stabilizing. Corrected reversals occur for Falcon in social Snowdrift and Harmony, Gemma in social Stag Hunt and Harmony, and Gemma in team Stag Hunt. Their concentration in social or team framings across several games shows that context can change both magnitude and direction. Including fallback outputs and partial trajectories changes none of the 96 corrected classifications (Supplementary Material, Section~A.3).

\section{Mechanistic Analysis}

The behavioral treatment changes messages and the histories they create. Experiments 1 and 2 instead alter current prompts along fixed communication-treatment histories, estimating immediate open-loop effects; Experiment 3 intervenes on a candidate component during regenerated play.

\begin{table*}[t]
\centering
\scriptsize
\setlength{\tabcolsep}{2.5pt}
\begin{tabular*}{\textwidth}{@{\extracolsep{\fill}}lcccc@{}}
\toprule
\textbf{Condition} & \textbf{Qwen} & \textbf{Falcon} & \textbf{Gemma} & \textbf{Granite} \\
\midrule
Partner message & \textbf{12.94 [12.19, 13.68]} & \textbf{4.40 [4.05, 4.75]} & \textbf{$-0.43$ [$-0.77$, $-0.10$]} & $-0.75$ [$-1.54$, 0.00] \\
Own text as partner & \textbf{13.13 [12.39, 13.88]} & \textbf{5.37 [4.96, 5.78]} & $-0.17$ [$-0.53$, 0.15] & 0.25 [$-0.56$, 1.00] \\
Partner text as own public & \textbf{4.38 [3.66, 5.09]} & \textbf{1.27 [0.98, 1.55]} & \textbf{$-0.50$ [$-0.77$, $-0.24$]} & \textbf{$-2.10$ [$-2.79$, $-1.42$]} \\
Own public message & \textbf{4.73 [4.04, 5.40]} & \textbf{1.59 [1.31, 1.88]} & $-0.27$ [$-0.54$, $-0.00$] & $-0.11$ [$-0.83$, 0.58] \\
Dummy text & \textbf{7.39 [6.67, 8.09]} & \textbf{4.82 [4.50, 5.14]} & \textbf{$-0.63$ [$-0.89$, $-0.37$]} & \textbf{2.42 [1.99, 2.87]} \\
Partner-formatted dummy & \textbf{$-8.19$ [$-9.08$, $-7.26$]} & \textbf{$-0.52$ [$-0.86$, $-0.17$]} & \textbf{$-6.38$ [$-6.77$, $-5.97$]} & \textbf{$-10.92$ [$-11.73$, $-10.10$]} \\
No message & 0.00 (reference) & 0.00 (reference) & 0.00 (reference) & 0.00 (reference) \\
\bottomrule
\end{tabular*}
\caption{Experiment 1 reduction in open-loop expected switching relative to no message (percentage points; 95\% intervals); positive values indicate persistence. Bold entries have Benjamini--Hochberg $q<.05$ across the six nonreference conditions within model. Estimates weight 24 game--context strata equally, from 10,000 paired bootstrap resamples of 20 dyads per stratum.}
\label{mech-policy}
\end{table*}

\subsection{Experiment 1: Current-Message Policy Replay}

Table~\ref{mech-policy} shows that Qwen has the clearest strategically specific immediate effect. Under matched partner framing, the endogenous message reduces expected switching by 21.12 points more than partner-formatted dummy text (95\% CI [19.99, 22.22]). Holding dummy content fixed, partner formatting changes the effect from 7.39 to $-8.19$ points, a 15.57-point loss [14.59, 16.53]; content and attribution therefore interact. Apparent source also matters: own text presented as partner speech yields 13.13 points, versus 4.38 when partner text is presented as the focal agent's message.

Falcon's partner-message and plain-dummy effects are similar (4.40 versus 4.82 points), and Gemma's partner effect is slightly destabilizing. In Granite, real partner content recovers 10.17 points [9.00, 11.29] relative to partner-formatted dummy text but remains near the no-message reference. Partner-formatted dummy text increases expected switching in every model, so partner formatting alone is insufficient. Immediate specificity is therefore strongest in Qwen; Granite's 73.5-point trajectory effect likely depends on accumulated interaction, earlier messages, self-generated anchoring, or their interaction rather than the latest received message alone.

\begin{figure}[t]
\centering
\includegraphics[width=0.97\columnwidth]{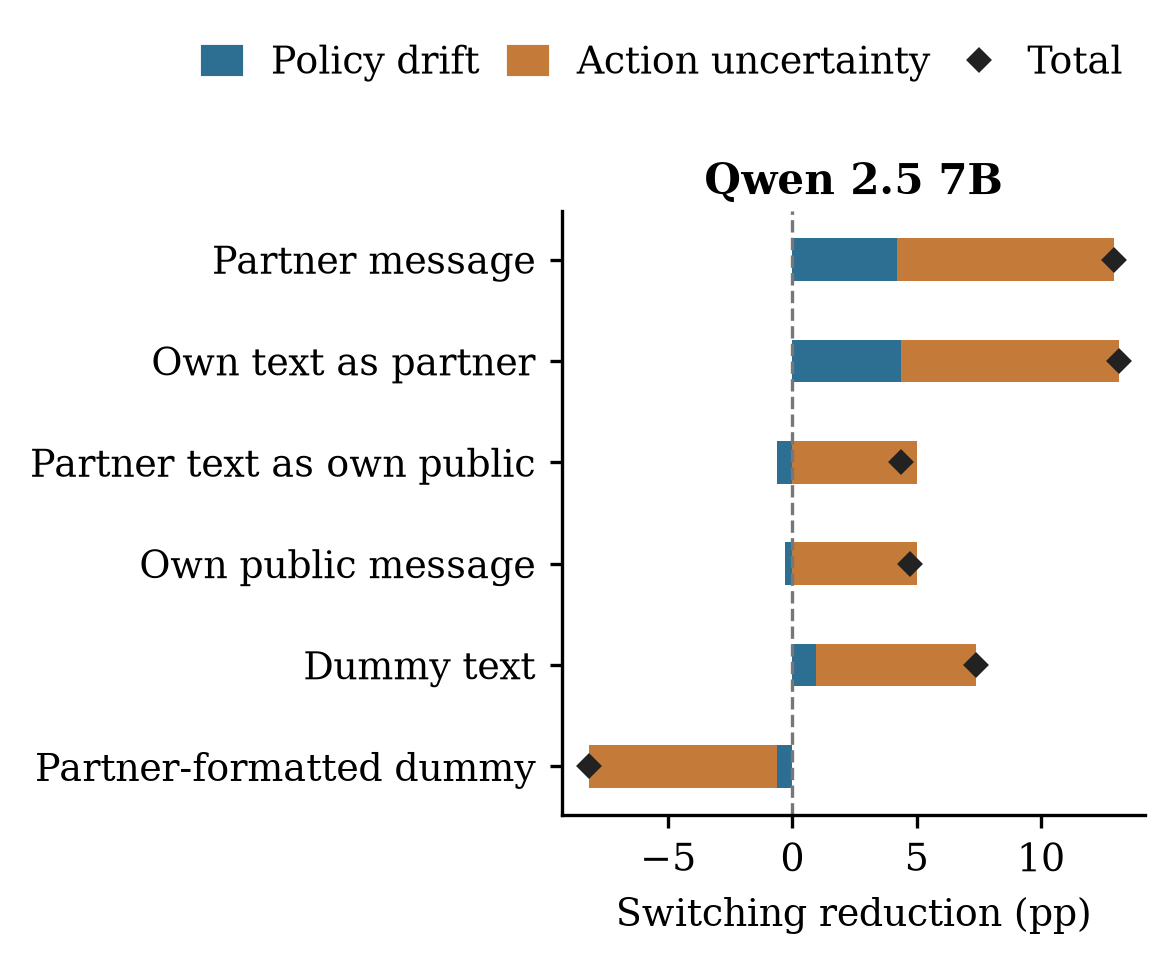}
\caption{Experiment 1 decomposition for Qwen. Bars partition each reduction in
expected switching into reduced between-round $P(A)$ drift (blue) and reduced
action uncertainty (orange); diamonds show their exact sum. Positive values
indicate greater persistence.}
\label{mech-decomposition}
\end{figure}

Figure~\ref{mech-decomposition} decomposes Qwen's partner-message effect into 4.22 points of reduced probability drift and 8.72 of reduced action uncertainty; dummy text yields 0.95 and 6.44 points. Generic text can sharpen the action distribution, whereas strategically attributed content additionally limits movement across rounds. Other-model decompositions appear in Supplementary Material, Figure~S2.

\subsection{Experiment 2: History-Conditioned Message Effects}

The pooled history-by-message interaction is 3.42 points for Qwen (95\% CI [3.06, 3.80]), 20.29 for Falcon [19.77, 20.81], 18.11 for Gemma [17.69, 18.53], and 16.31 for Granite [15.63, 16.98]; all remain significant after correction ($q<.001$). Conditional effects reverse sign: Gemma's SP-minus-MB contrast is 11.90 points after A and $-24.32$ after B; Granite's is 19.91 and $-12.70$. Averaging over recent behavior can therefore hide substantial responsiveness. Because Experiment 1 compares endogenous messages with deletion whereas Experiment 2 manipulates standardized content and history, the two are complementary rather than nested. Qwen's Snowdrift interaction is null; the other models are positive in every game (Supplementary Material, Section~C).

\begin{figure}[t]
\centering
\includegraphics[width=0.97\columnwidth]{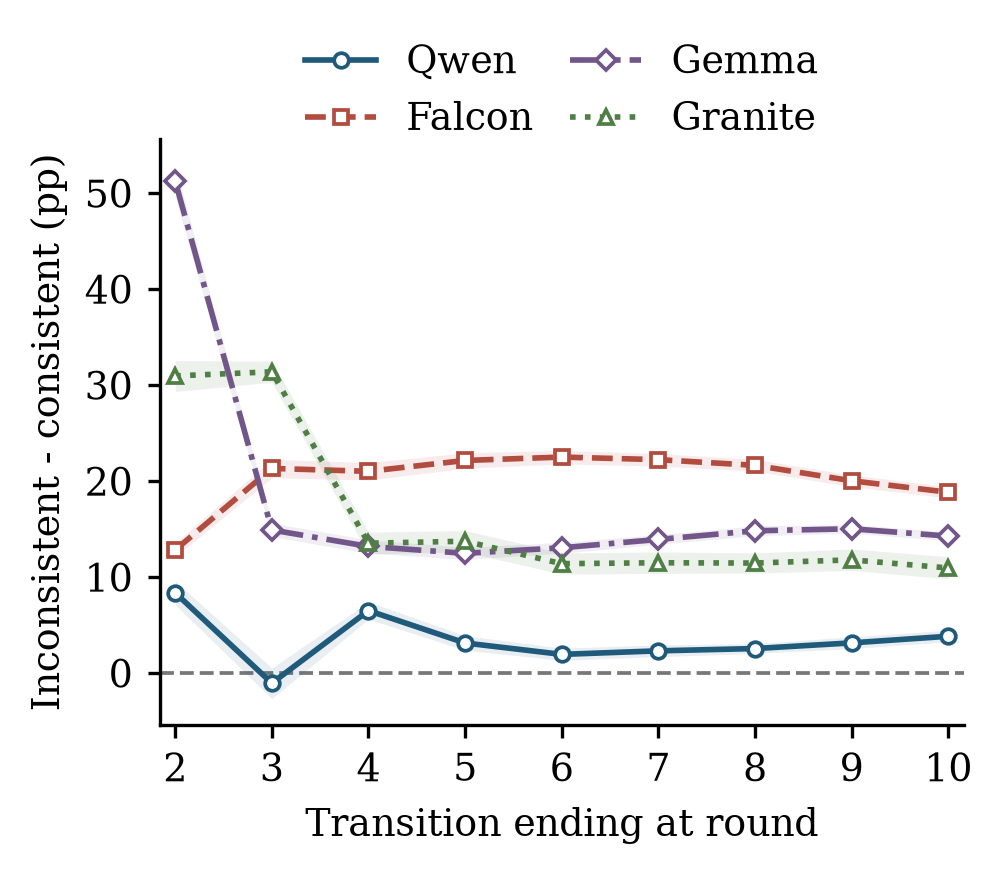}
\caption{Experiment 2 history-by-message interaction by transition. Points show the
one-half interaction defined in Methods; positive values mean the SP-minus-MB
contrast is larger after partner action A than after B. Estimates pool four
games and six contexts; bands are 95\% stratified-bootstrap intervals.}
\label{mech-history-round}
\end{figure}

Figure~\ref{mech-history-round} shows that conditioning appears early and persists rather than accumulating monotonically. Qwen is positive at eight of nine transitions; Falcon remains strongly positive and peaks mid-interaction; Gemma and Granite peak early and then settle to smaller positive effects. Current language is therefore interpreted against recent partner behavior from the beginning of play and throughout the interaction.

\subsection{Experiment 3: Layerwise Causal Intervention}

Adding the learned direction to held-out prompts gives small, sometimes sign-reversing A--B logit-margin effects at early layers. From about layer 18, both models show larger, consistently positive responses relative to equal-norm random directions.

Closed-loop projection produces no corrected increase in switching through layer 18, then significant increases at layers 20, 22, and 24 in both models. At layer 20, switching rises by 0.099 (95\% CI [0.042, 0.156]) in Qwen and 0.581 [0.487, 0.674] in Falcon. Qwen peaks at layer 22, rising from 6.0\% under clean play to 70.0\% under ablation ($\Delta=0.640$ [0.553, 0.727]). At Qwen layer 22, mean action entropy under ablation is only 0.125 nats, versus the binary maximum of 0.693; the intervention therefore induces confident alternation rather than near-random choice. It changes which action the model commits to across rounds rather than dissolving within-round commitment. Falcon remains significant at layer 26 ($\Delta=0.306$ [0.212, 0.401]); neither model is corrected-significant at layer 28. Because intervened actions enter subsequent histories, these are trajectory-level effects. The full sweep appears in Supplementary Material, Figure~S4.

Two limits qualify the causal claim. Equal-norm random controls were used for additive steering but not for closed-loop projection, so ablation establishes sensitivity to removing this component, not projection specificity. Moreover, mutual-benefit and self-prioritizing messages differ in favored action as well as semantics; \emph{policy-content direction} names the construction rather than a unique interpretation. The results therefore establish causal sensitivity of complete trajectories to a late-layer component, not unique mediation of communication-induced stabilization.

\section{Conclusion}

Across four repeated games, cheap talk reduces average adjacent-round switching from 63.8\% to 18.9\%, with 71 of 96 cell-level effects remaining stabilizing after correction. Effects span every incentive structure but vary sharply by model and context; five reversals occur in social or team framings.

Mechanistic analyses refine this aggregate pattern. In Qwen, strategic content and partner attribution jointly reduce action uncertainty and probability drift. Across all models, recent partner behavior conditions the effect of message framing, showing that messages are not interpreted in isolation. In Qwen and Falcon, removing a late-layer policy-content component increases switching during regenerated play, establishing causal relevance without projection specificity or unique mediation.

The study covers four open-weight 7--9B checkpoints, ten-round interactions, fixed payoff structures and framings, and temperature 0.8. Activation interventions are limited to depth-matched Qwen and Falcon in Prisoner's Dilemma with ten trajectories per context. Switching measures first-order action persistence rather than general predictability or strategic quality, and history-presentation robustness remains untested. The findings establish a replicated pattern within this model class; generalization to larger or reasoning-focused models, longer horizons, richer games, and alternative communication protocols remains open.

% Bibliography set slightly smaller (9.6/10.7 instead of 10/11) so the
% reference list ends at the foot of page 9.
{\fontsize{9.6}{10.7}\selectfont
\bibliography{references}
}

% ===========================================================================
% ======================  SUPPLEMENTARY MATERIAL  ===========================
% ===========================================================================

\clearpage
\setcounter{secnumdepth}{2}
\appendix
\setcounter{section}{0}
\setcounter{figure}{0}
\setcounter{table}{0}
\setcounter{equation}{0}
\renewcommand{\thetable}{S\arabic{table}}
\renewcommand{\thefigure}{S\arabic{figure}}
\renewcommand{\theequation}{S\arabic{equation}}
\raggedbottom
\emergencystretch=1em

\twocolumn[{%
  \begin{center}
    \rule{\textwidth}{1.2pt}\\[0.9em]
    {\Large\bfseries SUPPLEMENTARY MATERIAL}\\[0.5em]
    {\large Cheap Talk Stabilizes Strategic Interaction in LLM Agents}\\[0.7em]
    {\normalsize End of main text. Sections, figures, tables, and equations
     below are numbered A, B, C\ldots and S1, S2, S3\ldots}\\[0.6em]
    \rule{\textwidth}{1.2pt}\\[1.2em]
  \end{center}%
}]

\section{Scope and Analysis Inventory}

This supplement provides additional behavioral evidence, mechanistic analyses, and implementation details supporting the main paper. It first expands the primary behavioral results, then reports the three mechanistic experiments, and finally documents the computational and prompt-level provenance of the study.

\smallskip
\noindent\textbf{Behavioral experiment.}
Four models play four ten-round games in six contexts, with 100 dyads in every model--game--context--treatment cell. The primary outcome is realized adjacent-round action switching.

\smallskip
\noindent\textbf{Experiment 1: current-message policy replay.}
For 20 communication-treatment dyads per model--game--context cell, 38,400 action decisions are rescored under seven current-message conditions. The outcome is policy-implied expected switching along fixed realized histories.

\smallskip
\noindent\textbf{Experiment 2: history-conditioned message effects.}
The same decisions are rescored under a $2\times2$ crossing of recent partner action and standardized current-message framing. The outcome is a history-by-message interaction in expected switching.

\smallskip
\noindent\textbf{Replay reference.}
Each decision is additionally scored once under \texttt{pre\_action}, which replays the unmodified action prompt from the actual accumulated conversation state. This pass is used only as a natural-policy reference for reconstruction checks and is not an Experiment 1 or Experiment 2 condition.

\smallskip
\noindent\textbf{Experiment 3: layerwise causal intervention.}
Qwen and Falcon, the only depth-matched pair among the four models, are tested in Prisoner's Dilemma across six contexts at layers $2,4,\ldots,28$. One-step tests use 150 held-out no-message action decisions, balanced at 25 per context; closed-loop tests compare 60 paired clean and ablated trajectories at each layer.

\subsection{Behavioral Outcomes and Payoffs}

Reduced switching does not by itself reveal which action becomes persistent or whether stabilization improves realized outcomes. We therefore return to the 100-dyad behavioral archive and summarize marginal action frequencies, joint outcomes, and realized payoffs. For each game and treatment, we first compute the relevant quantity within every model--context cell and then give the resulting 24 cells equal weight. Confidence intervals use 10,000 dyad-level bootstrap resamples within each cell. Communication and no-communication dyads are resampled independently because they are separate trajectories.

\begin{table*}[t]
\centering
\scriptsize
\setlength{\tabcolsep}{3.2pt}
\begin{tabular*}{\textwidth}{@{\extracolsep{\fill}}lrrrrrr@{}}
\toprule
& \multicolumn{3}{c}{\textbf{Action-A rate (\%)}} & \multicolumn{3}{c}{\textbf{Payoff per agent per round}} \\
\cmidrule(lr){2-4}\cmidrule(lr){5-7}
\textbf{Game} & \textbf{No message} & \textbf{Message} & \textbf{$\Delta$ [95\% CI]} & \textbf{No message} & \textbf{Message} & \textbf{$\Delta$ [95\% CI]} \\
\midrule
Prisoner's Dilemma & 58.56 & 82.05 & 23.49 [22.36, 24.64] & 2.335 & 2.739 & 0.404 [0.384, 0.423] \\
Snowdrift           & 61.57 & 84.44 & 22.87 [21.88, 23.84] & 2.246 & 2.808 & 0.563 [0.538, 0.587] \\
Stag Hunt           & 70.64 & 89.13 & 18.49 [17.50, 19.44] & 3.382 & 4.366 & 0.984 [0.925, 1.044] \\
Harmony             & 70.49 & 89.90 & 19.41 [18.57, 20.27] & 3.395 & 4.431 & 1.036 [0.988, 1.085] \\
\bottomrule
\end{tabular*}
\caption{Action-A frequencies and realized payoffs by game and treatment. Differences are communication minus no communication. Action A is reported descriptively rather than assigned a common strategic label across games. Estimates give equal weight to 24 model--context cells and use only complete, error-free dyads.}
\label{supp-behavior-outcomes}
\end{table*}

\begin{table*}[t]
\centering
\scriptsize
\setlength{\tabcolsep}{2.7pt}
\begin{tabular*}{\textwidth}{@{\extracolsep{\fill}}llrrrrrr@{}}
\toprule
& & \multicolumn{3}{c}{\textbf{Action-A rate (\%)}} & \multicolumn{3}{c}{\textbf{Payoff per agent per round}} \\
\cmidrule(lr){3-5}\cmidrule(lr){6-8}
\textbf{Model} & \textbf{Game} & \textbf{No msg.} & \textbf{Msg.} & \textbf{$\Delta$ [95\% CI]} & \textbf{No msg.} & \textbf{Msg.} & \textbf{$\Delta$ [95\% CI]} \\
\midrule
Qwen & Prisoner's Dilemma & 43.72 & 90.54 & 46.83 [45.07, 48.50] & 1.978 & 2.866 & 0.888 [0.853, 0.922] \\
& Snowdrift & 50.01 & 92.15 & 42.14 [40.99, 43.23] & 1.696 & 2.916 & 1.220 [1.180, 1.259] \\
& Stag Hunt & 47.94 & 92.36 & 44.42 [43.00, 45.84] & 2.610 & 4.566 & 1.956 [1.874, 2.038] \\
& Harmony & 50.10 & 94.37 & 44.27 [43.29, 45.19] & 2.370 & 4.682 & 2.312 [2.254, 2.369] \\
\addlinespace
Falcon & Prisoner's Dilemma & 70.47 & 79.76 & 9.29 [7.09, 11.51] & 2.612 & 2.736 & 0.124 [0.091, 0.157] \\
& Snowdrift & 77.00 & 83.02 & 6.02 [3.92, 8.16] & 2.806 & 2.865 & 0.059 [0.024, 0.095] \\
& Stag Hunt & 86.51 & 86.15 & $-0.36$ [$-2.31$, 1.62] & 4.119 & 4.137 & 0.018 [$-0.105$, 0.144] \\
& Harmony & 85.54 & 86.22 & 0.67 [$-1.28$, 2.58] & 4.166 & 4.205 & 0.040 [$-0.073$, 0.152] \\
\addlinespace
Granite & Prisoner's Dilemma & 48.30 & 75.62 & 27.32 [25.08, 29.42] & 2.170 & 2.630 & 0.460 [0.419, 0.501] \\
& Snowdrift & 49.57 & 80.56 & 31.00 [28.96, 32.96] & 1.842 & 2.715 & 0.873 [0.809, 0.936] \\
& Stag Hunt & 57.22 & 85.68 & 28.47 [26.14, 30.76] & 2.404 & 4.217 & 1.813 [1.675, 1.944] \\
& Harmony & 53.18 & 85.03 & 31.84 [29.77, 33.92] & 2.453 & 4.179 & 1.726 [1.613, 1.839] \\
\addlinespace
Gemma & Prisoner's Dilemma & 71.75 & 82.29 & 10.54 [7.72, 13.26] & 2.579 & 2.724 & 0.144 [0.099, 0.190] \\
& Snowdrift & 69.70 & 82.01 & 12.31 [9.99, 14.65] & 2.638 & 2.735 & 0.098 [0.048, 0.147] \\
& Stag Hunt & 90.89 & 92.33 & 1.44 [$-0.52$, 3.37] & 4.396 & 4.546 & 0.149 [0.026, 0.274] \\
& Harmony & 93.13 & 94.01 & 0.87 [$-0.72$, 2.48] & 4.591 & 4.656 & 0.065 [$-0.029$, 0.161] \\
\bottomrule
\end{tabular*}
\caption{Model--game expansion of Table~\ref{supp-behavior-outcomes}. Within each model--game combination, estimates give equal weight to the six contexts.}
\label{supp-behavior-outcomes-model-game}
\end{table*}

\begin{table}[t]
\centering
\scriptsize
\setlength{\tabcolsep}{3.0pt}
\begin{tabular}{llrrrr}
\toprule
\textbf{Game} & \textbf{Treatment} & \textbf{AA} & \textbf{AB} & \textbf{BA} & \textbf{BB} \\
\midrule
Prisoner's Dilemma & No message & 42.21 & 16.01 & 16.70 & 25.09 \\
& Message & 72.28 & 9.91 & 9.63 & 8.18 \\
Snowdrift & No message & 48.29 & 13.25 & 13.31 & 25.15 \\
& Message & 75.27 & 9.06 & 9.27 & 6.40 \\
Stag Hunt & No message & 55.87 & 15.25 & 14.30 & 14.59 \\
& Message & 82.51 & 6.52 & 6.73 & 4.24 \\
Harmony & No message & 57.53 & 13.10 & 12.82 & 16.56 \\
& Message & 83.44 & 6.58 & 6.35 & 3.63 \\
\bottomrule
\end{tabular}
\caption{Joint-action frequencies (\%) under the same equal-cell weighting as Table~\ref{supp-behavior-outcomes}. In the asymmetric outcomes AB and BA, the first letter denotes Agent 0's action and the second denotes Agent 1's.}
\label{supp-joint-outcomes}
\end{table}

\newpage
Across games, communication raises the marginal frequency of Action A by 18.49--23.49 percentage points and increases average payoff by 0.404--1.036 points per agent per round. The joint-outcome shift is similarly systematic: AA increases by approximately 25.9--30.1 points, while AB, BA, and BB all decline. Stabilization is therefore accompanied by a pronounced shift toward persistent AA play and higher realized scores rather than arbitrary persistence in either action. The magnitude of this shift remains model-dependent. Qwen and Granite move strongly toward A in every game, whereas Falcon and Gemma change little in Stag Hunt and Harmony, where their no-communication Action-A rates are already high.

\subsection{Period-Two Cycling}

For an agent action sequence $a_{id1},\ldots,a_{idT}$, define period-two adherence as
\begin{equation}
C_{id}=\frac{1}{T-2}\sum_{t=3}^{T}
\mathbf{1}\!\left\{a_{idt}=a_{id,t-2}\ \text{and}\ a_{idt}\neq a_{id,t-1}\right\}.
\label{eq:period-two}
\end{equation}
A score of one means that every action from round 3 onward repeats the action taken two rounds earlier while differing from the immediately preceding action. The measure therefore distinguishes systematic alternation from a high switch rate produced by irregular, nonperiodic changes.

\begin{figure*}[t]
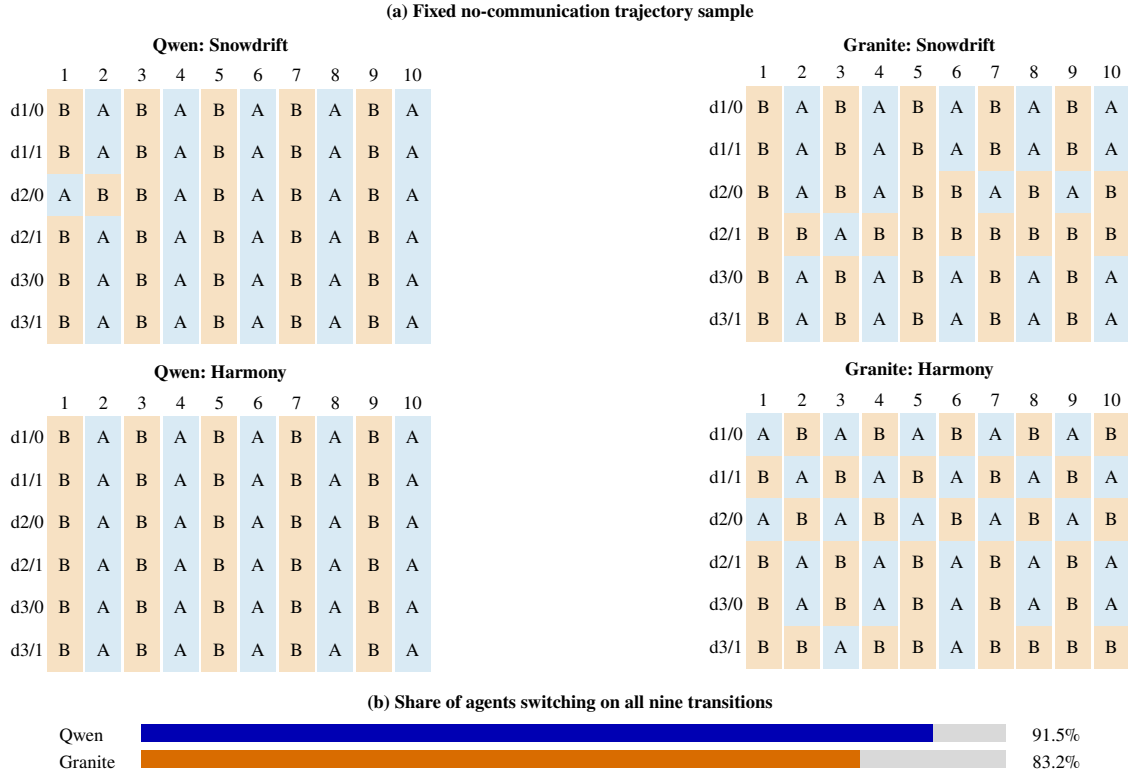

\centering
\scriptsize
\textbf{(a) Fixed no-communication trajectory sample}\par\smallskip
\begin{minipage}[t]{0.48\textwidth}
\centering
\textbf{Qwen: Snowdrift}\par\vspace{2pt}
\setlength{\tabcolsep}{0.6pt}
\begin{tabular}{l*{10}{c}}
&1&2&3&4&5&6&7&8&9&10\\
d1/0&\trajB&\trajA&\trajB&\trajA&\trajB&\trajA&\trajB&\trajA&\trajB&\trajA\\
d1/1&\trajB&\trajA&\trajB&\trajA&\trajB&\trajA&\trajB&\trajA&\trajB&\trajA\\
d2/0&\trajA&\trajB&\trajB&\trajA&\trajB&\trajA&\trajB&\trajA&\trajB&\trajA\\
d2/1&\trajB&\trajA&\trajB&\trajA&\trajB&\trajA&\trajB&\trajA&\trajB&\trajA\\
d3/0&\trajB&\trajA&\trajB&\trajA&\trajB&\trajA&\trajB&\trajA&\trajB&\trajA\\
d3/1&\trajB&\trajA&\trajB&\trajA&\trajB&\trajA&\trajB&\trajA&\trajB&\trajA\\
\end{tabular}

\vspace{7pt}\textbf{Qwen: Harmony}\par\vspace{2pt}
\begin{tabular}{l*{10}{c}}
&1&2&3&4&5&6&7&8&9&10\\
d1/0&\trajB&\trajA&\trajB&\trajA&\trajB&\trajA&\trajB&\trajA&\trajB&\trajA\\
d1/1&\trajB&\trajA&\trajB&\trajA&\trajB&\trajA&\trajB&\trajA&\trajB&\trajA\\
d2/0&\trajB&\trajA&\trajB&\trajA&\trajB&\trajA&\trajB&\trajA&\trajB&\trajA\\
d2/1&\trajB&\trajA&\trajB&\trajA&\trajB&\trajA&\trajB&\trajA&\trajB&\trajA\\
d3/0&\trajB&\trajA&\trajB&\trajA&\trajB&\trajA&\trajB&\trajA&\trajB&\trajA\\
d3/1&\trajB&\trajA&\trajB&\trajA&\trajB&\trajA&\trajB&\trajA&\trajB&\trajA\\
\end{tabular}
\end{minipage}\hfill
\begin{minipage}[t]{0.48\textwidth}
\centering
\textbf{Granite: Snowdrift}\par\vspace{2pt}
\setlength{\tabcolsep}{0.6pt}
\begin{tabular}{l*{10}{c}}
&1&2&3&4&5&6&7&8&9&10\\
d1/0&\trajB&\trajA&\trajB&\trajA&\trajB&\trajA&\trajB&\trajA&\trajB&\trajA\\
d1/1&\trajB&\trajA&\trajB&\trajA&\trajB&\trajA&\trajB&\trajA&\trajB&\trajA\\
d2/0&\trajB&\trajA&\trajB&\trajA&\trajB&\trajB&\trajA&\trajB&\trajA&\trajB\\
d2/1&\trajB&\trajB&\trajA&\trajB&\trajB&\trajB&\trajB&\trajB&\trajB&\trajB\\
d3/0&\trajB&\trajA&\trajB&\trajA&\trajB&\trajA&\trajB&\trajA&\trajB&\trajA\\
d3/1&\trajB&\trajA&\trajB&\trajA&\trajB&\trajA&\trajB&\trajA&\trajB&\trajA\\
\end{tabular}

\vspace{7pt}\textbf{Granite: Harmony}\par\vspace{2pt}
\begin{tabular}{l*{10}{c}}
&1&2&3&4&5&6&7&8&9&10\\
d1/0&\trajA&\trajB&\trajA&\trajB&\trajA&\trajB&\trajA&\trajB&\trajA&\trajB\\
d1/1&\trajB&\trajA&\trajB&\trajA&\trajB&\trajA&\trajB&\trajA&\trajB&\trajA\\
d2/0&\trajA&\trajB&\trajA&\trajB&\trajA&\trajB&\trajA&\trajB&\trajA&\trajB\\
d2/1&\trajB&\trajA&\trajB&\trajA&\trajB&\trajA&\trajB&\trajA&\trajB&\trajA\\
d3/0&\trajB&\trajA&\trajB&\trajA&\trajB&\trajA&\trajB&\trajA&\trajB&\trajA\\
d3/1&\trajB&\trajB&\trajA&\trajB&\trajB&\trajA&\trajB&\trajB&\trajB&\trajB\\
\end{tabular}
\end{minipage}

\vspace{8pt}
\textbf{(b) Share of agents switching on all nine transitions}\par\smallskip
\setlength{\tabcolsep}{5pt}
\begin{tabular}{@{}l l r@{}}
Qwen
& \textcolor{blue!70!black}{\rule{4.12in}{2.3mm}}\textcolor{black!15}{\rule{0.38in}{2.3mm}}
& 91.5\% \\
Granite
& \textcolor{orange!85!black}{\rule{3.74in}{2.3mm}}\textcolor{black!15}{\rule{0.76in}{2.3mm}}
& 83.2\% \\
\end{tabular}
\caption{Near-period-two behavior without communication. Panel (a) shows both agents from the first three numerically indexed complete, error-free dyads in each neutral-context cell; trajectories were not selected by outcome. Blue cells denote A and orange cells denote B. Panel (b) pools all four games and six contexts within each model and reports the share of agents that switch on all nine adjacent-round transitions.}
\label{supp-period-two-figure}
\end{figure*}

\begin{table*}[!t]
\centering
\scriptsize
\setlength{\tabcolsep}{2.5pt}
\begin{tabular*}{\textwidth}{@{\extracolsep{\fill}}llrrrrr@{}}
\toprule
& & \multicolumn{3}{c}{\textbf{Agent trajectories}} & \multicolumn{2}{c}{\textbf{Dyads}} \\
\cmidrule(lr){3-5}\cmidrule(lr){6-7}
\textbf{Model} & \textbf{Game} & \textbf{9/9 switches} & \textbf{$\geq$8/9 switches} & \textbf{$C_{id}$, median [IQR]} & \textbf{Both agents 9/9} & \textbf{In phase $\,|\, $both 9/9} \\
\midrule
Qwen & Prisoner's Dilemma & 80.3\% & 81.4\% & 1.00 [1.00, 1.00] & --- & --- \\
& Snowdrift & 98.5\% & 98.8\% & 1.00 [1.00, 1.00] & 583/600 (97.2\%) & 514/583 (88.2\%) \\
& Stag Hunt & 87.9\% & 89.4\% & 1.00 [1.00, 1.00] & --- & --- \\
& Harmony & 99.3\% & 99.5\% & 1.00 [1.00, 1.00] & 591/600 (98.5\%) & 434/591 (73.4\%) \\
\addlinespace
Granite & Prisoner's Dilemma & 86.0\% & 91.8\% & 1.00 [1.00, 1.00] & --- & --- \\
& Snowdrift & 92.9\% & 97.5\% & 1.00 [1.00, 1.00] & 526/599 (87.8\%) & 419/526 (79.7\%) \\
& Stag Hunt & 66.0\% & 77.8\% & 1.00 [0.75, 1.00] & --- & --- \\
& Harmony & 87.9\% & 92.3\% & 1.00 [1.00, 1.00] & 498/600 (83.0\%) & 294/498 (59.0\%) \\
\bottomrule
\end{tabular*}
\caption{No-communication alternation in Qwen and Granite, pooling the six contexts within each game. Each complete, error-free dyad contributes two agent trajectories. Dyad-level phase is reported for Snowdrift and Harmony: in-phase dyads cycle between AA and BB, whereas the complementary out-of-phase dyads cycle between AB and BA. One error-marked Granite--Snowdrift dyad is excluded, leaving 599 dyads and 1,198 agent trajectories; every other cell contains 600 dyads and 1,200 agent trajectories.}
\label{supp-period-two-table}
\end{table*}

Pooling all four games, 91.5\% of Qwen agents and 83.2\% of Granite agents switch on all nine transitions; 92.3\% and 89.8\%, respectively, switch at least eight times. Median period-two adherence equals one in every model--game cell. The extreme no-communication switch rates therefore reflect a population-wide alternating pattern rather than a small number of anomalous trajectories or an artifact of population averaging. In Harmony, both agents switch on all nine transitions in 591/600 Qwen and 498/600 Granite dyads; 434/591 (73.4\%) and 294/498 (59.0\%) of these are in phase. In Snowdrift, both agents alternate perfectly in 583/600 Qwen and 526/599 complete Granite dyads; 514/583 (88.2\%) and 419/526 (79.7\%) are in phase. The antiphase Snowdrift cycles that tie constant A at $(T+S)/2=R$ therefore account for only 69/583 (11.8\%) and 107/526 (20.3\%) of perfectly alternating dyads.

\newpage
\subsection{Error Audit}
\label{supp-error-audit-section}

\begin{table*}[t]
\centering
\scriptsize
\setlength{\tabcolsep}{2.5pt}
\begin{tabularx}{\textwidth}{@{}p{0.55in}p{0.72in}YYYYYY@{}}
\toprule
\textbf{Model} & \textbf{Treatment} & \textbf{Planned dyads} & \textbf{Error-marked dyads} & \textbf{Malformed-output events} & \textbf{API-failure events} & \textbf{Partial dyads} & \textbf{Complete clean dyads} \\
\midrule
Qwen & No message & 2,400 & 0 & 0 & 0 & 0 & 2,400 \\
& Message & 2,400 & 0 & 0 & 0 & 0 & 2,400 \\
Falcon & No message & 2,400 & 11 & 11 & 0 & 6 & 2,389 \\
& Message & 2,400 & 14 & 14 & 0 & 8 & 2,386 \\
Granite & No message & 2,400 & 1 & 1 & 0 & 0 & 2,399 \\
& Message & 2,400 & 12 & 15 & 0 & 10 & 2,388 \\
Gemma & No message & 2,400 & 1 & 1 & 0 & 1 & 2,399 \\
& Message & 2,400 & 0 & 0 & 0 & 0 & 2,400 \\
\bottomrule
\end{tabularx}
\caption{Behavioral error audit across four games and six contexts. The 42 malformed-output events comprise 41 unparsable action responses and one empty message. No archived trajectory records an Ollama/API exception, and a dyad may contain more than one malformed event.}
\label{supp-error-audit}
\end{table*}

The primary behavioral experiment planned 19,200 dyads. Thirty-nine dyads (0.20\%) contain at least one malformed output, including 25 that terminate before round ten. The complete-case analysis retains 19,161 dyads and 344,898 agent-level adjacent-round transitions. Mean switch rates are 0.6377 without communication and 0.1889 with communication, a reduction of 0.4488. Of the 96 model--game--context contrasts, 81 are positive; after Benjamini--Hochberg correction, 71 are stabilizing and five are destabilizing. Including fallback B actions and every available transition from partial trajectories yields rates of 0.6375 and 0.1890, a reduction of 0.4485. Under this specification, 82 contrasts are positive, and none of the 96 corrected significance classifications changes.

\subsection{Policy Extraction}

At each realized action decision, the replay runner sends the exact system prompt, conversation state, and counterfactual action prompt to the same Ollama checkpoint that generated the trajectory. Generation is limited to one token, with temperature 0.8, top-$p=1$, an 8,192-token context window, and 20 requested first-token log probabilities. If $\ell_{A,t}$ and $\ell_{B,t}$ denote the returned log probabilities of action tokens A and B at round $t$, the renormalized binary action policy is
\begin{equation}
p_t \equiv P_t(A)
=\frac{\exp(\ell_{A,t})}
{\exp(\ell_{A,t})+\exp(\ell_{B,t})}.
\label{eq:binary-policy}
\end{equation}
The corresponding probability of B is $P_t(B)=1-p_t$. Renormalization over A and B isolates the model's relative preference between the two valid actions while excluding probability mass assigned to formatting or explanatory tokens.

The exported archive contains 38,400 unique agent--round decisions and 12 scored passes per decision. Seven passes are the Experiment 1 current-message conditions, four are the Experiment 2 history--message conditions, and the remaining \texttt{pre\_action} pass replays the unmodified action prompt using the actual accumulated conversation state. The \texttt{pre\_action} pass serves only as a natural-policy reference for reconstruction checks and is not included in the Experiment 1 or Experiment 2 estimands. Each pass contains exactly 38,400 rows, so
\[
38{,}400 + 268{,}800 + 153{,}600 = 460{,}800,
\]
where the three terms correspond to \texttt{pre\_action}, the seven Experiment 1 passes, and the four Experiment 2 passes, respectively.

When one action token falls outside the returned top-20 window, its log probability is known only to be no greater than the lowest returned score. This yields an identification interval for $p_t$. The raw export records the interval, and the analysis uses the endpoint nearest $P_t(A)=0.5$, which is conservative with respect to policy extremity. Across the 460,800-row replay archive, 2,699 rows (0.59\%) are interval-censored: 0 for Qwen, 177 for Falcon, 29 for Gemma, and 2,493 for Granite. Eleven API responses without usable log probabilities were discarded and successfully retried. Nine incomplete behavioral attempts were likewise logged, discarded, and replaced before policy scoring. All 96 model--game--context cells contain the planned 20 complete dyads.

\subsection{Pairing and Inference}

Experiments 1 and 2 pair output-policy contrasts within model, dyad, agent, round, game, and context. Decision-level measurements are first aggregated within dyad. The bootstrap then resamples the 20 dyads with replacement inside each fixed game--context stratum and gives the 24 strata equal weight within a model. Confidence intervals and $p$-values use 10,000 paired stratified bootstrap resamples. In Experiment 3, one-step slopes are averaged within 25 context--simulation groups and evaluated with one-sample Student-$t$ intervals and tests; closed-loop inference uses the same procedure on 60 paired ablated-minus-clean trajectory differences per layer. Benjamini--Hochberg adjustment is applied separately to the prespecified testing families: the six condition-versus-no-message tests and five specificity contrasts within each model in Experiment 1; the four pooled model tests and, separately, the 16 model--game tests in Experiment 2; and the fourteen tested layers within each model and outcome in Experiment 3. Unless otherwise stated, corrected significance is defined by $q<.05$.

\FloatBarrier
\section{Experiment 1: Current-Message Policy Replay}

\subsection{Counterfactual Construction}

Experiment 1 asks which properties of the text immediately preceding an action make the next-action policy more persistent. For every action decision in a realized communication-treatment trajectory, we construct seven prompts that share the system prompt, game, context, focal agent, round, action instruction, and preceding realized history. Only the current text block and its displayed speaker role vary. Table~\ref{supp-message-conditions} gives the exact construction.

\begin{table*}[!t]
\centering
\footnotesize
\renewcommand{\arraystretch}{1.04}
\begin{tabularx}{\textwidth}{@{}p{1.30in}p{3.55in}X@{}}
\toprule
\textbf{Condition} & \textbf{Current block before the action instruction} & \textbf{What it isolates} \\
\midrule
Partner message
& ``Message received from counterpart: Agent $j$: [endogenous partner text]''
& Natural received communication \\
Own text as partner
& ``Message received from counterpart: Agent $j$: [endogenous focal-agent text]''
& Partner attribution with different endogenous content \\
Partner text as own public
& The endogenous partner text is displayed as a public message sent earlier by the focal agent
& Identical partner content without partner attribution \\
Own public message
& The endogenous focal-agent text is displayed as its own earlier public message
& Self-generated strategic content without received text \\
Dummy text
& ``Extra unrelated text shown before the action phase: [fixed dummy text]''
& Generic text without strategic content \\
Partner-formatted dummy
& ``Message received from counterpart: Agent $j$: [fixed dummy text]''
& Partner formatting without strategic content \\
No message
& The current text block is deleted
& History-only reference \\
\bottomrule
\end{tabularx}
\caption{Experiment 1 counterfactual conditions. Only the current text block changes. Endogenous text is drawn from the same focal agent, partner, and realized round. The fixed dummy text is ``The sky is blue. Grass is green. Seven is odd. Squares have four sides.''}
\label{supp-message-conditions}
\end{table*}

The conditions separate three features that coincide during natural play: strategically relevant content, true authorship, and displayed conversational role. The two cross-attribution conditions are particularly informative. ``Own text as partner'' preserves partner attribution while changing the underlying endogenous message and its true author. ``Partner text as own public'' holds the partner's exact text fixed while removing partner attribution.

\subsection{Expected Policy Switching and Decomposition}

Let $p_t^{(c)}$ denote the counterfactual $P_t(A)$ at round $t$ under condition $c$. Assuming independent draws from adjacent action policies, the probability that the draws disagree is
\begin{equation}
\begin{aligned}
E_t^{(c)}
&=p_t^{(c)}\!\left(1-p_{t+1}^{(c)}\right) \\
&\quad+\left(1-p_t^{(c)}\right)p_{t+1}^{(c)}.
\end{aligned}
\label{eq:expected-switch}
\end{equation}
This quantity is invariant to relabeling A and B and is low for both persistent-A and persistent-B policies.

Expected switching decomposes exactly as
\begin{equation}
\begin{aligned}
E_t^{(c)}
&=\underbrace{\left(p_{t+1}^{(c)}-p_t^{(c)}\right)^2}_{\text{probability drift}} \\
&\quad+\underbrace{p_t^{(c)}\!\left(1-p_t^{(c)}\right)
+p_{t+1}^{(c)}\!\left(1-p_{t+1}^{(c)}\right)}_{\text{action uncertainty}}.
\end{aligned}
\label{eq:switch-decomposition}
\end{equation}
The first component is the squared between-round change in $P(A)$ and therefore measures movement in the model's relative preference for A versus B. The second is the sum of the two adjacent policies' Bernoulli variances. Expected switching can therefore arise from probability drift, within-round action uncertainty, or both. For each condition, we report
\begin{equation}
\Delta S_c
=\overline{E}_{\mathrm{no\ message}}-\overline{E}_c,
\label{eq:condition-effect}
\end{equation}
so positive values indicate that condition $c$ lowers the policy-implied probability of switching relative to the no-message replay.

\subsection{Results and Strategic Specificity}

The main paper reports the complete condition estimates, and Figure~\ref{supp-mech-conditions} displays them jointly. Qwen shows the clearest strategically specific pattern. Under matched partner framing, its endogenous partner message is 21.12 percentage points [19.99, 22.22] more stabilizing than the dummy message. Under matched dummy content, plain presentation is 15.57 points [14.59, 16.53] more stabilizing than partner-attributed presentation. Presenting the focal agent's endogenous message as partner speech yields a 13.13-point reduction, whereas presenting the partner's exact message as the focal agent's own public text reduces the effect to 4.38 points. The Qwen result therefore reflects a strong interaction between content and apparent source rather than a main effect of either alone.

\begin{table}[t]
\centering
\scriptsize
\setlength{\tabcolsep}{2.5pt}
\begin{tabularx}{\columnwidth}{@{}lYY@{}}
\toprule
\textbf{Model} & \textbf{Real vs. dummy, partner framed} & \textbf{Plain vs. partner-framed dummy} \\
\midrule
Qwen
& \textbf{21.12 [19.99, 22.22]}
& \textbf{15.57 [14.59, 16.53]} \\
Falcon
& \textbf{4.93 [4.51, 5.32]}
& \textbf{5.34 [5.07, 5.62]} \\
Gemma
& \textbf{5.95 [5.44, 6.45]}
& \textbf{5.75 [5.31, 6.17]} \\
Granite
& \textbf{10.17 [9.00, 11.29]}
& \textbf{13.33 [12.53, 14.14]} \\
\bottomrule
\end{tabularx}
\caption{Matched Experiment 1 contrasts in percentage points with 95\% paired-bootstrap confidence intervals. The first column changes content while retaining partner framing; the second changes framing while retaining identical dummy content. Positive values favor the first condition named in each heading. Bold intervals exclude zero.}
\label{supp-specificity}
\end{table}

Both matched contrasts are positive in every model. Real content improves persistence relative to irrelevant content under the partner wrapper, while attaching that wrapper to irrelevant content reduces persistence relative to plain presentation. Partner framing is therefore consequential but not generically stabilizing: its effect depends on what the attributed partner actually says. Qwen and Falcon combine this matched content advantage with a positive endogenous partner-message effect relative to deleting the message.

\begin{figure}[!b]
\centering
\includegraphics[width=\columnwidth]{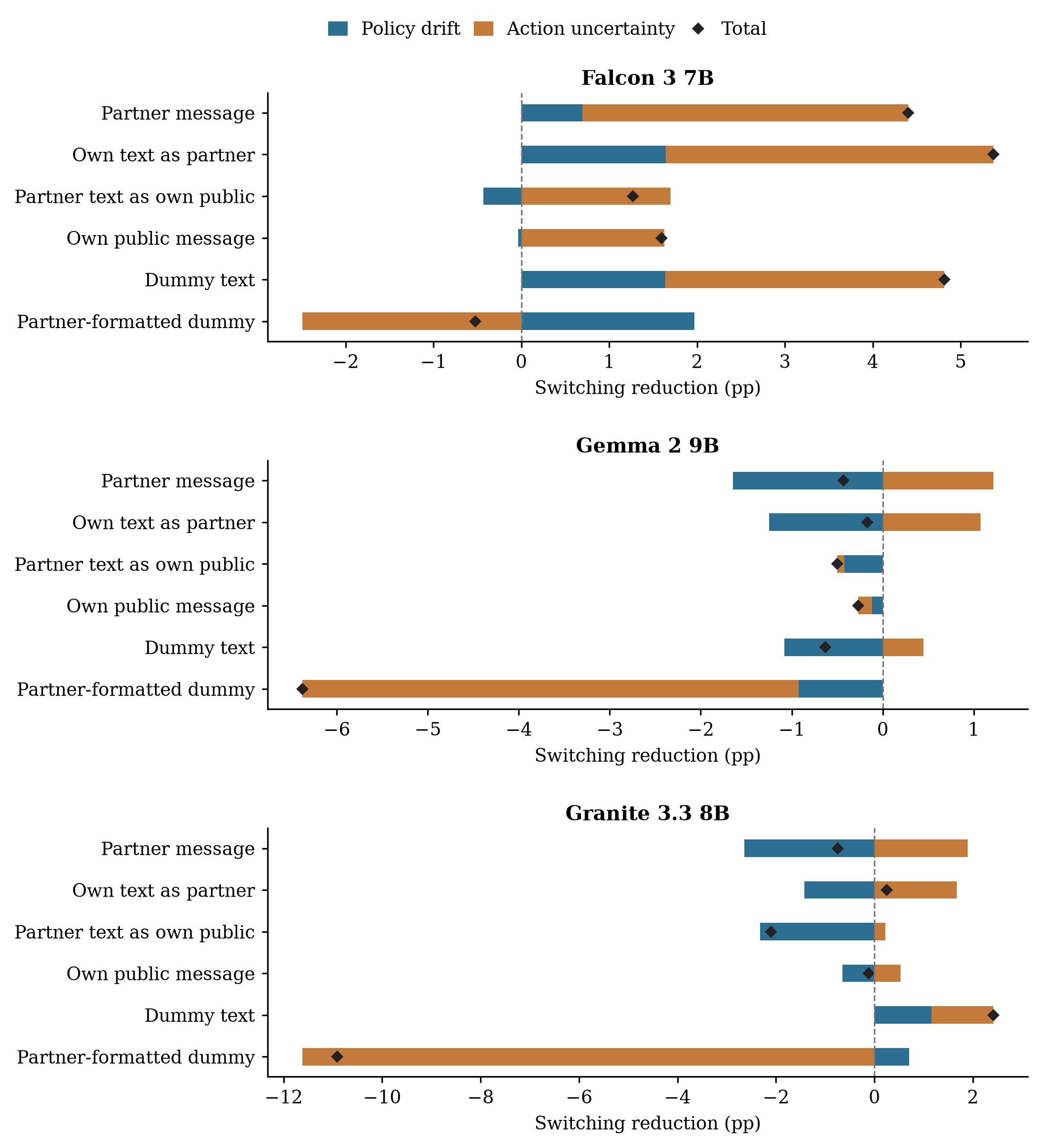}
\caption{Experiment 1 decomposition for Falcon, Gemma, and Granite. Bars partition the no-message-minus-condition effect into reduced between-round $P(A)$ drift (blue) and reduced action uncertainty (orange); diamonds show their exact sum. Positive totals indicate lower expected switching. The corresponding Qwen decomposition appears in the main paper.}
\label{supp-mech-decomposition-other}
\end{figure}

Figure~\ref{supp-mech-decomposition-other} reports the decomposition for Falcon, Gemma, and Granite; the corresponding Qwen panel appears in the main paper. For Qwen's endogenous partner message, 4.22 points of the 12.94-point total come from reduced between-round probability drift and 8.72 points from reduced action uncertainty. Dummy text produces only 0.95 points of drift reduction but 6.44 points of uncertainty reduction. Generic text can therefore sharpen Qwen's action policy, whereas partner-attributed strategic content additionally limits movement in action probabilities across rounds. Falcon's positive effects are dominated by lower uncertainty. In Gemma and Granite, the endogenous partner message reduces uncertainty but increases probability drift enough to eliminate a positive total effect.

\FloatBarrier
\section{Experiment 2: History-Conditioned Message Effects}

\subsection{Matched History--Message Design}

Experiment 2 asks whether recent partner behavior conditions how current message framing affects policy persistence. The focal agent's preceding action is fixed at A, the partner's preceding action is set to A or B, and the current partner message expresses either mutual benefit or self-prioritization. These manipulations produce the complete $2\times2$ factorial in Table~\ref{supp-history-conditions}.

\begin{center}
\begin{minipage}{\columnwidth}
\centering
\footnotesize
\setlength{\tabcolsep}{2.5pt}
\begin{tabularx}{\columnwidth}{@{}p{0.45in}Xp{0.55in}@{}}
\toprule
\textbf{Latest partner action} & \textbf{Current partner message} & \textbf{Cell} \\
\midrule
A
& ``I intend to pursue the mutually beneficial outcome this round.''
& A/MB \\
A
& ``I intend to prioritize my own highest possible score this round.''
& A/SP \\
B
& ``I intend to pursue the mutually beneficial outcome this round.''
& B/MB \\
B
& ``I intend to prioritize my own highest possible score this round.''
& B/SP \\
\bottomrule
\end{tabularx}
\makeatletter
\def\@captype{table}
\makeatother
\caption{Experiment 2 factorial conditions. The partner's latest action is crossed with current message framing; MB denotes mutual-benefit language and SP denotes self-prioritizing language.}
\label{supp-history-conditions}
\end{minipage}
\end{center}

From round 2 onward, the synthetic outcome block replaces the natural preceding-round summary at every scored decision. At round 1, an analogous synthetic prior-outcome block is inserted so that the factorial manipulation is defined at every round index. All remaining inputs---including model, game, context, focal identity, round index, earlier history, and action instruction---remain paired.

\subsection{History-by-Message Estimand}

Let $\overline{E}_{A,\mathrm{MB}}$ denote mean expected switching when the partner's latest action is A and the current message expresses mutual benefit. Define the other three cells analogously, with SP denoting self-prioritization. The symmetrically scaled history-by-message interaction is
\begin{equation}
\Delta S_{\mathrm{HM}}
=\frac{1}{2}\left[
\left(\overline{E}_{A,\mathrm{SP}}-\overline{E}_{A,\mathrm{MB}}\right)
-\left(\overline{E}_{B,\mathrm{SP}}-\overline{E}_{B,\mathrm{MB}}\right)
\right].
\label{eq:history-effect}
\end{equation}
This is one-half of the conventional $2\times2$ difference-in-differences and retains an average-difference scale. A nonzero value means that the message-framing effect depends on recent partner action.

\subsection{Pooled and Game-Specific Results}

\begin{table*}[t]
\centering
\scriptsize
\setlength{\tabcolsep}{3.6pt}
\begin{tabular*}{\textwidth}{@{\extracolsep{\fill}}lccccc@{}}
\toprule
\textbf{Model} & \textbf{Pooled} & \textbf{Prisoner's Dilemma} & \textbf{Snowdrift} & \textbf{Stag Hunt} & \textbf{Harmony} \\
\midrule
Qwen 2.5 7B
& \textbf{3.42 [3.06, 3.80]}
& \textbf{1.15 [0.51, 1.85]}
& $-0.40$ [$-1.16$, 0.36]
& \textbf{6.73 [6.14, 7.36]}
& \textbf{6.21 [5.38, 7.09]} \\
Falcon 3 7B
& \textbf{20.29 [19.77, 20.81]}
& \textbf{20.49 [19.44, 21.52]}
& \textbf{21.67 [20.66, 22.68]}
& \textbf{19.55 [18.53, 20.57]}
& \textbf{19.45 [18.34, 20.52]} \\
Gemma 2 9B
& \textbf{18.11 [17.69, 18.53]}
& \textbf{20.47 [19.74, 21.20]}
& \textbf{19.33 [18.48, 20.17]}
& \textbf{18.99 [18.04, 19.93]}
& \textbf{13.65 [12.81, 14.44]} \\
Granite 3.3 8B
& \textbf{16.31 [15.63, 16.98]}
& \textbf{14.82 [13.31, 16.37]}
& \textbf{15.69 [14.24, 17.11]}
& \textbf{16.95 [15.70, 18.17]}
& \textbf{17.77 [16.54, 19.03]} \\
\bottomrule
\end{tabular*}
\caption{History-by-message interaction $\Delta S_{\mathrm{HM}}$ in percentage points with 95\% confidence intervals. Pooled estimates give equal weight to 24 game--context strata; game-specific estimates give equal weight to six contexts. Bold pooled entries have $q<.05$ after correction across the four model tests, and bold game-specific entries have $q<.05$ after correction across the 16 model--game tests.}
\label{supp-history-summary}
\end{table*}

The pooled interaction is positive in all four models. Its conditional contrasts can have opposite signs: the self-prioritizing-minus-mutual-benefit expected-switching contrast is 3.20 points after partner action A and $-3.64$ after B for Qwen, 39.79 and $-0.80$ for Falcon, 11.90 and $-24.32$ for Gemma, and 19.91 and $-12.70$ for Granite. Averaging over histories can therefore conceal strong conditional responsiveness. Qwen's 3.42-point interaction is more heterogeneous: Snowdrift is not distinguishable from zero, while its other game estimates remain positive after correction. Falcon, Gemma, and Granite are positive in all four games.

\subsection{Components and Round Profile}

\begin{figure}[t]
\centering
\includegraphics[width=\columnwidth]{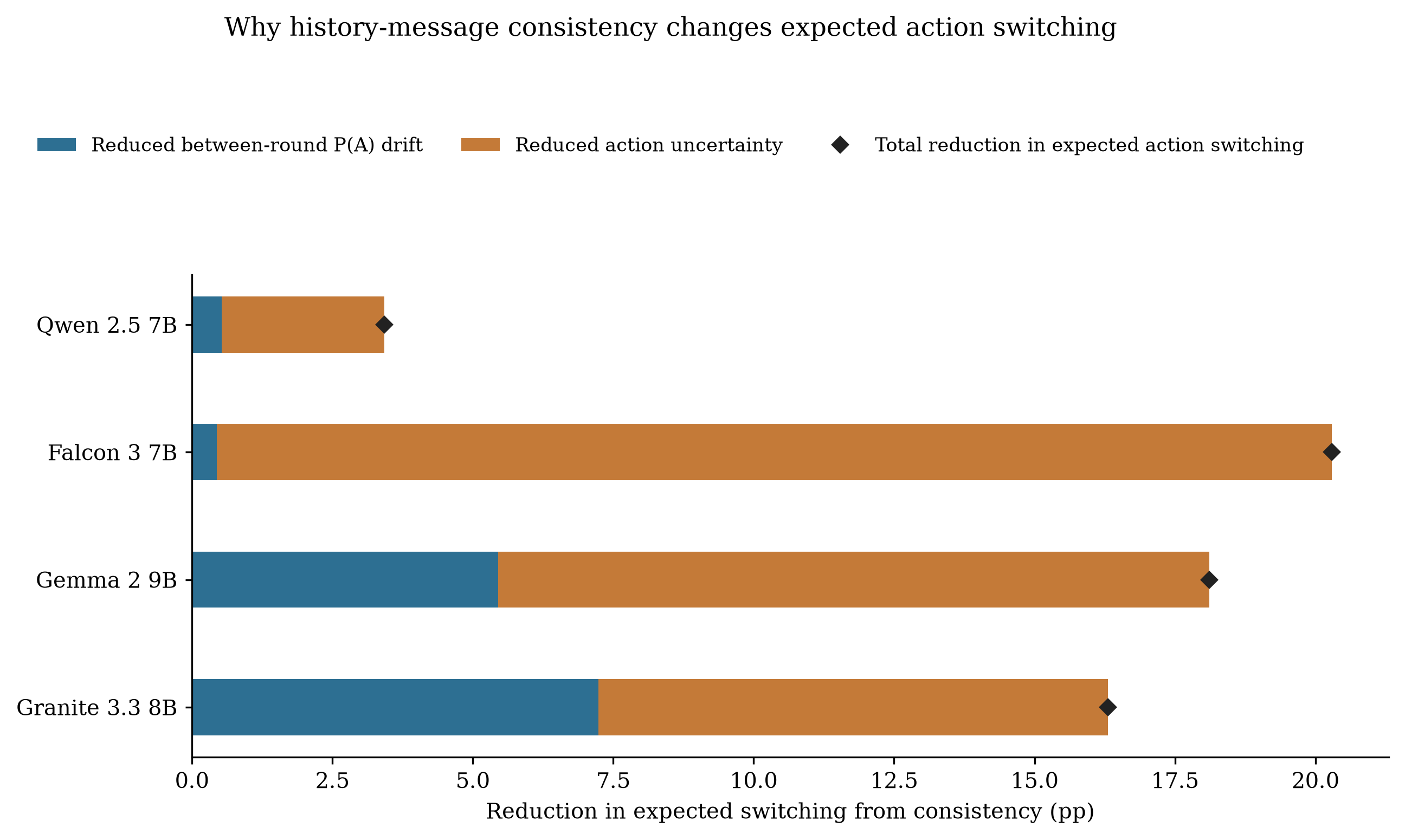}
\caption{Exact decomposition of the Experiment 2 history-by-message interaction. Blue bars show the probability-drift component, orange bars show the action-uncertainty component, and diamonds show their total.}
\label{supp-history-decomposition}
\end{figure}

Figure~\ref{supp-history-decomposition} partitions the pooled interaction using Equation~\ref{eq:switch-decomposition}. Drift and uncertainty contribute 0.53 and 2.89 points for Qwen, 0.45 and 19.84 for Falcon, 5.45 and 12.66 for Gemma, and 7.24 and 9.07 for Granite. Action uncertainty is the dominant common component; probability drift contributes more in Gemma and Granite than in Qwen and Falcon.

The main-paper round profile covers transitions ending at rounds 2--10. Qwen is positive at eight of nine transitions but unresolved at round 3; Falcon remains strongly positive; Gemma and Granite are largest early and then settle to smaller positive values. The interaction therefore appears early and persists, with model-dependent timing and magnitude.

\FloatBarrier
\section{Experiment 3: Layerwise Causal Intervention}

\begin{figure*}[!t]
\centering
\includegraphics[width=0.94\textwidth]{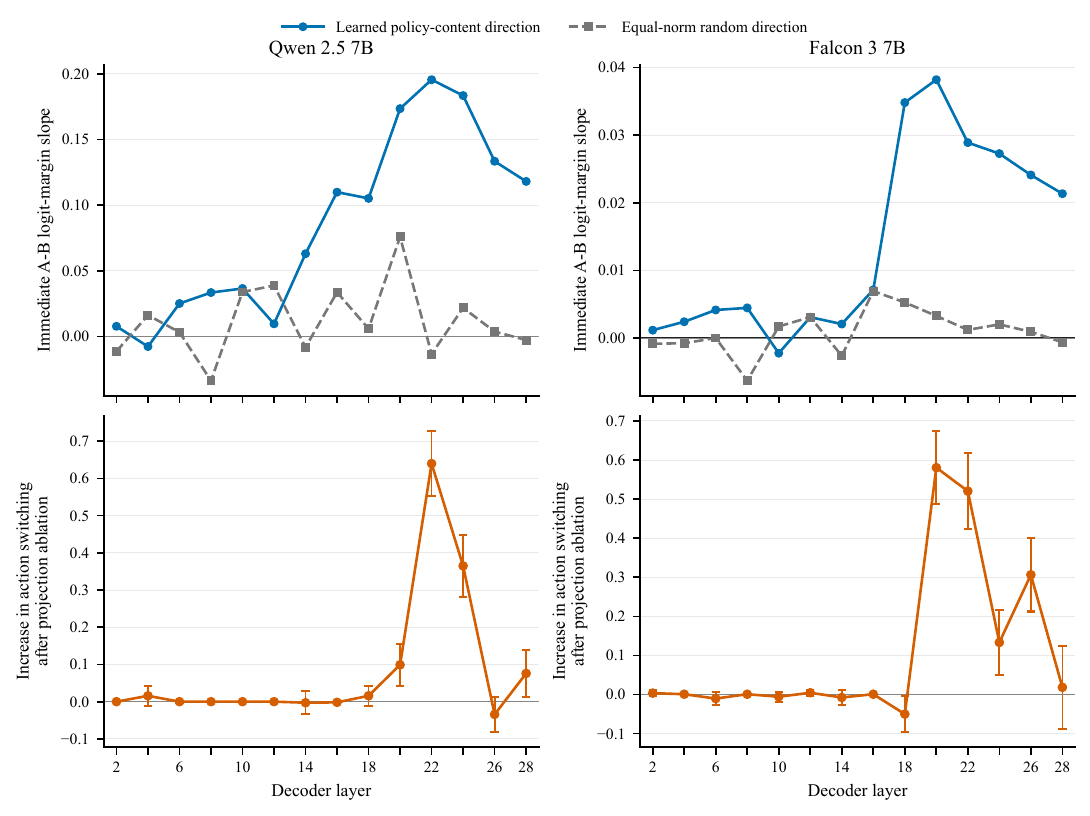}
\caption{Experiment 3 layerwise interventions in the depth-matched Qwen and Falcon models across decoder layers 2--28. Top: immediate A--B logit-margin slopes for the learned direction (blue) and an equal-norm random direction (gray). Bottom: paired change in realized ten-round action switching after projection ablation, with 95\% confidence intervals. Positive bottom-panel values indicate that ablation increases switching. The closed-loop analysis uses 60 context--seed trajectory pairs per layer.}
\label{supp-mech-layerwise}
\end{figure*}

\subsection{Depth-Matched Activation Capture}

The causal experiment uses Qwen 2.5 7B Instruct and Falcon 3 7B Instruct, the only depth-matched pair among the four evaluated models. Both contain 28 transformer layers, whereas Granite and Gemma contain 40 and 42, respectively. Matching total depth permits an identical intervention grid, $2,4,\ldots,28$, and comparison at the same absolute layer indices and fractions of network depth without introducing model depth as an additional difference. The restriction is therefore a controlled design choice and does not imply that the candidate mechanism is absent from Granite or Gemma.

The same Ollama weights used in the behavioral simulations can be loaded through the Transformers GGUF stack. The source models are \texttt{qwen2.5:7b} and \texttt{falcon3:7b}, both using Q4\_K\_M quantization. For activation capture, the tensors are loaded into the corresponding \path{Qwen/Qwen2.5-7B-Instruct} and \path{tiiuae/Falcon3-7B-Instruct} architectures and dequantized to \texttt{float16}.

For each model, the capture cohort contains ten Prisoner's Dilemma communication-treatment trajectories in each of the six contexts. The 1,200 complete decision keys are split deterministically into 780 construction and 420 held-out keys (65\%/35\%), ensuring that no decision both defines and evaluates the intervention direction.

\subsection{History-Balanced Policy Direction}

At each layer $l$, we collect final-token residual-stream activations for the four Experiment 2 variants: the mutual-benefit message under synthetic partner-A and partner-B histories, and the self-prioritizing message under the same two histories. Each of the 780 construction decisions contributes two activation rows to each message class, yielding 1,560 source rows per class and balancing partner-A and partner-B frames exactly. If $\mu_l^{\mathrm{MB}}$ and $\mu_l^{\mathrm{SP}}$ denote the two class means, the candidate direction is
\begin{equation}
d_l
=\frac{\mu_l^{\mathrm{MB}}-\mu_l^{\mathrm{SP}}}
{\left\lVert\mu_l^{\mathrm{MB}}-\mu_l^{\mathrm{SP}}\right\rVert_2}.
\label{eq:direction}
\end{equation}
Balancing partner-A and partner-B histories within each message class prevents the mean difference from being driven by unequal representation of the partner's latest action.

\subsection{Immediate and Closed-Loop Tests}

The one-step test adds the candidate direction at the final prompt position:
\begin{equation}
h'_{l,-1}=h_{l,-1}+\alpha d_l,
\label{eq:addition-intervention}
\end{equation}
For $\alpha\in\{-2,-1,-0.5,0.5,1,2\}$, we estimate the slope of the A--B logit margin, $\ell_A-\ell_B$, with respect to $\alpha$. Evaluation uses 150 held-out \texttt{cf\_no\_message} decisions, selected as 25 per context. The same decisions are perturbed with an equal-norm random direction. The reported immediate effect is the learned-direction slope minus the random-direction slope, controlling for the generic effect of a same-norm residual-stream perturbation.

The stronger test removes the candidate component during every action choice in a regenerated ten-round trajectory:
\begin{equation}
h'_{l,-1}
=h_{l,-1}-\left(h_{l,-1}^{\top}d_l\right)d_l.
\label{eq:projection-ablation}
\end{equation}
Message generation remains unchanged. Each intervened trajectory is paired with a clean communication-treatment trajectory using the same model, context, and seed. The closed-loop outcome is $S_{\mathrm{ablated}}-S_{\mathrm{clean}}$, where $S$ denotes realized adjacent-round switching. A positive value means that removing the component makes the regenerated trajectory less persistent. The equal-norm random comparison belongs only to the additive test; no matched random-projection trajectory is included.

\FloatBarrier

\begin{table*}[t]
\centering
\scriptsize
\setlength{\tabcolsep}{3.4pt}
\begin{tabular*}{\textwidth}{@{\extracolsep{\fill}}rcccc@{}}
\toprule
& \multicolumn{2}{c}{\textbf{Qwen 2.5 7B}} & \multicolumn{2}{c}{\textbf{Falcon 3 7B}} \\
\cmidrule(lr){2-3}\cmidrule(lr){4-5}
\textbf{Layer}
& \textbf{Learned $-$ random slope}
& \textbf{Ablated $-$ clean switching}
& \textbf{Learned $-$ random slope}
& \textbf{Ablated $-$ clean switching} \\
\midrule
2  & 0.019 [$-0.003$, 0.041] & 0.000 [0.000, 0.000] & \textbf{0.002 [0.001, 0.003]} & 0.004 [$-0.004$, 0.011] \\
4  & \textbf{$-0.024$ [$-0.031$, $-0.017$]} & 0.016 [$-0.011$, 0.042] & \textbf{0.003 [0.003, 0.004]} & 0.001 [$-0.001$, 0.003] \\
6  & \textbf{0.022 [0.017, 0.027]} & 0.000 [$-0.003$, 0.003] & \textbf{0.004 [0.003, 0.005]} & $-0.010$ [$-0.028$, 0.007] \\
8  & \textbf{0.067 [0.057, 0.077]} & 0.000 [0.000, 0.000] & \textbf{0.011 [0.009, 0.012]} & 0.001 [$-0.001$, 0.003] \\
10 & 0.003 [$-0.013$, 0.018] & 0.000 [0.000, 0.000] & \textbf{$-0.004$ [$-0.005$, $-0.003$]} & $-0.006$ [$-0.018$, 0.007] \\
12 & \textbf{$-0.029$ [$-0.042$, $-0.017$]} & 0.000 [0.000, 0.000] & 0.000 [$-0.001$, 0.001] & 0.005 [$-0.003$, 0.013] \\
14 & \textbf{0.071 [0.059, 0.083]} & $-0.003$ [$-0.034$, 0.028] & \textbf{0.005 [0.003, 0.007]} & $-0.007$ [$-0.026$, 0.011] \\
16 & \textbf{0.076 [0.044, 0.109]} & $-0.002$ [$-0.006$, 0.002] & 0.000 [$-0.001$, 0.001] & 0.001 [$-0.001$, 0.003] \\
18 & \textbf{0.099 [0.067, 0.132]} & 0.016 [$-0.011$, 0.043] & \textbf{0.030 [0.027, 0.032]} & $-0.050$ [$-0.096$, $-0.004$] \\
20 & \textbf{0.098 [0.082, 0.114]} & \textbf{0.099 [0.042, 0.156]} & \textbf{0.035 [0.032, 0.037]} & \textbf{0.581 [0.487, 0.674]} \\
22 & \textbf{0.209 [0.190, 0.227]} & \textbf{0.640 [0.553, 0.727]} & \textbf{0.028 [0.026, 0.030]} & \textbf{0.520 [0.424, 0.617]} \\
24 & \textbf{0.162 [0.154, 0.169]} & \textbf{0.365 [0.282, 0.448]} & \textbf{0.025 [0.024, 0.027]} & \textbf{0.133 [0.049, 0.217]} \\
26 & \textbf{0.130 [0.127, 0.133]} & $-0.034$ [$-0.082$, 0.013] & \textbf{0.023 [0.022, 0.024]} & \textbf{0.306 [0.212, 0.401]} \\
28 & \textbf{0.121 [0.119, 0.123]} & 0.076 [0.012, 0.140] & \textbf{0.022 [0.021, 0.023]} & 0.019 [$-0.087$, 0.124] \\
\bottomrule
\end{tabular*}
\caption{Complete Experiment 3 layer sweep with 95\% Student-$t$ confidence intervals. The one-step columns report the learned-minus-random A--B logit-margin slope across 25 context--simulation means. The closed-loop columns report the ablated-minus-clean change in switch-rate proportion across 60 paired context--seed trajectories. Bold entries have $q<.05$ after separate Benjamini--Hochberg correction across fourteen layers within each model and outcome.}
\label{supp-layer-table}
\end{table*}

\begin{figure*}[!t]
\centering
\includegraphics[width=0.80\textwidth]{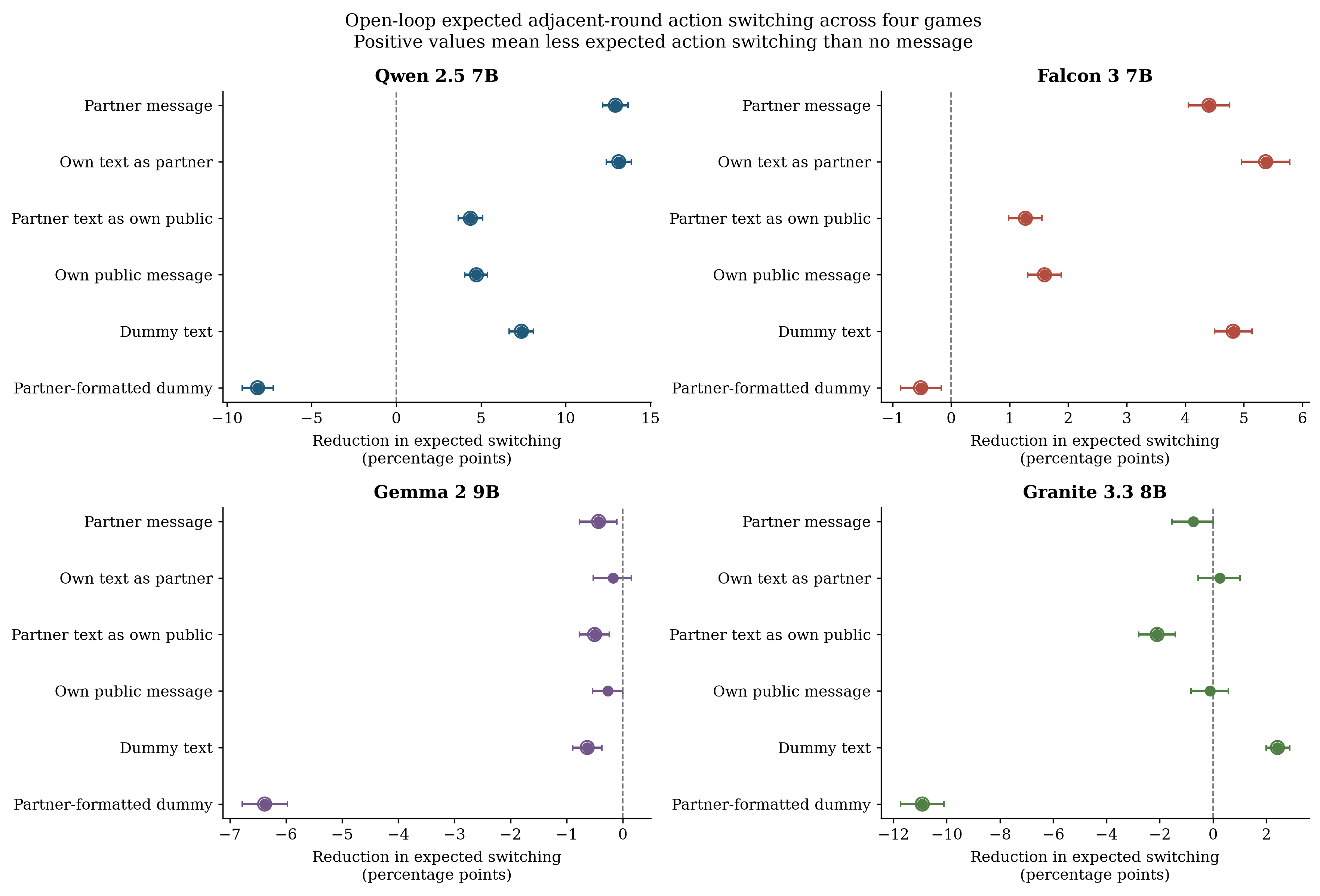}
\caption{Complete Experiment 1 condition display. Each point is the reduction in open-loop expected adjacent-round switching relative to no message, averaged with equal weight across four games and six contexts. Positive values indicate greater policy persistence; no message is zero by construction. Error bars are 95\% paired stratified-bootstrap confidence intervals based on 20 dyads per stratum.}
\label{supp-mech-conditions}
\end{figure*}

\subsection{Layerwise Interpretation}

Immediate A--B logit-margin sensitivity appears at several early layers, but the effects are small and their signs vary. The closed-loop result is more localized: projection ablation produces no Benjamini--Hochberg-significant increase in switching through layer 18, then significantly increases switching at layers 20, 22, and 24 in both models and at layer 26 in Falcon. Qwen peaks at layer 22 ($\Delta S=0.640$ [0.553, 0.727]), whereas Falcon peaks at layer 20 ($\Delta S=0.581$ [0.487, 0.674]) and remains positive at layer 26 ($\Delta S=0.306$ [0.212, 0.401]). Neither model has a corrected effect at layer 28. On the tested grid, the clearest trajectory-level dependence on the history-balanced policy content component therefore lies in a model-dependent late-layer band spanning layers 20--26.

All Qwen projection-ablation trajectories are error-free. Falcon records parser fallbacks in one of 60 trajectories at layers 6, 20, 24, and 26 and in two trajectories at layer 22; the layer-26 trajectory contains six fallback actions, while each other affected trajectory contains one. Excluding every affected pair leaves the corrected closed-loop conclusions unchanged: layers 20, 22, 24, and 26 remain significant.

\section{Additional Experiment 1 Display}

Figure~\ref{supp-mech-conditions} collects the complete set of current-message replay estimates in one panel, complementing the numerical results in the main paper and the specificity contrasts reported above.
\FloatBarrier

\section{Reproducibility Details}

\subsection{Code and Data Organization}

The released code follows the same stages as the study. Run notebooks are in \path{notebooks/}; matching analysis scripts are in \path{analysis/}. Each notebook writes the raw outputs for its stage and can resume from saved checkpoints.

\begin{center}
\scriptsize
\setlength{\tabcolsep}{3pt}
\begin{tabularx}{\columnwidth}{@{}p{0.58in}X@{}}
\toprule
\textbf{Stage} & \textbf{Run notebook and analysis script} \\
\midrule
Behavioral
& \path{00_run_behavioral_ollama.ipynb}\newline
  \path{00_analyze_behavioral_switching.py} \\
Experiment 1
& \path{01_run_message_and_history_counterfactuals.ipynb}\newline
  \path{01_analyze_message_conditions.py} \\
Experiment 2
& Same run notebook as Experiment 1\newline
  \path{02_analyze_history_vetting.py} \\
Experiment 3
& \path{02_run_layerwise_causal_interventions.ipynb}\newline
  \path{03_analyze_layerwise_interventions.py} \\
\bottomrule
\end{tabularx}
\end{center}

The behavioral stage records ten-round trajectories, and \path{00_analyze_behavioral_switching.py} produces the switching summaries; the action-frequency, payoff, and period-two results in Section~A are direct aggregations of the same trajectory archive. Experiments 1 and 2 record first-token A/B policy scores and produce the message-condition and history-by-message estimates. Experiment 3 records activations and intervention outcomes and produces the layerwise causal results. Shared functions are provided under \path{src/cheaptalk/}; \path{requirements-analysis.txt} and \path{requirements-experiment3.txt} define the released environments.

\subsection{Models and Software}

\begin{center}
\scriptsize
\setlength{\tabcolsep}{2.5pt}
\begin{tabularx}{\columnwidth}{@{}p{0.72in}p{0.82in}p{0.52in}X@{}}
\toprule
\textbf{Model} & \textbf{Ollama name} & \textbf{Quant.} & \textbf{Study stages} \\
\midrule
Qwen 2.5 7B & \texttt{qwen2.5:7b} & Q4\_K\_M & Behavioral; Exp. 1--3 \\
Falcon 3 7B & \texttt{falcon3:7b} & Q4\_K\_M & Behavioral; Exp. 1--3 \\
Granite 3.3 8B & \texttt{granite3.3:8b} & Q4\_K\_M & Behavioral; Exp. 1--2 \\
Gemma 2 9B & \texttt{gemma2:9b} & Q4\_0 & Behavioral; Exp. 1--2 \\
\bottomrule
\end{tabularx}
\end{center}

All behavioral and policy-replay calls use an 8,192-token context window. Experiment 3 reloads the Qwen and Falcon GGUF weights into their corresponding Transformers architectures, as described in the Experiment 3 methods. The original behavioral runs did not retain a complete package and GPU snapshot. The released requirements files define the environment for the final analyses and Experiment 3 implementation.

\subsection{Decoding and Randomness}

\noindent\textbf{Behavioral trajectories.}
Temperature is 0.8, top-$p$ is 1.0, and the context window is 8,192 tokens. Generation length uses the runtime default; \texttt{top\_k}, \texttt{min\_p}, repetition penalty, and caller-level stop strings are unset. The prompt requests messages of at most 20 tokens, but the API does not enforce that limit programmatically.

\smallskip
\noindent\textbf{Experiments 1--2.}
Policy replay uses the same temperature, top-$p$, and context window, generating one token with \texttt{top\_logprobs=20}. A call is retried up to twice if log probabilities or an A/B candidate are missing.

\smallskip
\noindent\textbf{Experiment 3.}
Activation capture and closed-loop interventions use temperature 0.8, top-$p$ 1.0, and a 2,048-token Ollama-matched context budget, with generation limits of 64 message tokens and 8 action tokens. The matched GGUF weights are dequantized to \texttt{float16}; centroid-replacement evaluations generate one action token.

\noindent\textbf{Model-template stops.}
Falcon adds \texttt{<|system|>}, \texttt{<|user|>}, \texttt{<|end|>}, and \texttt{<|assistant|>}; Gemma adds \texttt{<start\_of\_turn>} and \texttt{<end\_of\_turn>}. Qwen and Granite add no model-specific stop strings.

\smallskip
\noindent\textbf{Behavioral randomness.}
The primary runner does not pass an explicit seed to Ollama. Simulation indices denote separate stochastic calls, and the communication and no-communication trajectories are independently generated. Bootstrap resampling uses seed 42.

\smallskip
\noindent\textbf{Mechanistic randomness.}
The Experiment 3 runner seeds Python, NumPy, PyTorch, and all CUDA devices. Capture trajectory $i$ uses seed $42+i$. Event keys are shuffled with seed 20,260,701 and split into 65\% direction-construction and 35\% held-out sets. The layer-$l$ random control uses seed $20{,}260{,}701+99+l$. Closed-loop trajectory $i$ uses seed $100{,}042+i$; each clean and ablated pair shares its model, context, and seed, yielding 60 paired trajectories per model and layer.

\subsection{Message-Exchange Protocol}

The two message calls execute sequentially at the API level, but the first message is stored only in a local buffer and is not inserted into the second agent's history. Both buffered messages are revealed only after the two agents have finished composing. Actions are likewise generated without revealing the counterpart's current action. The implemented information flow is:

\begin{quote}\footnotesize\ttfamily
for each round t:\\
\hspace*{1em}generate Agent 0 message\\
\hspace*{2em}(history through t-1 only)\\
\hspace*{1em}buffer Agent 0 message\\
\hspace*{1em}generate Agent 1 message\\
\hspace*{2em}(history through t-1 only)\\
\hspace*{1em}buffer Agent 1 message\\[2pt]
\hspace*{1em}reveal Agent 1 message to Agent 0\\
\hspace*{1em}reveal Agent 0 message to Agent 1\\
\hspace*{1em}generate Agent 0 action\\
\hspace*{2em}(counterpart action hidden)\\
\hspace*{1em}generate Agent 1 action\\
\hspace*{2em}(counterpart action hidden)\\
\hspace*{1em}score joint action; update histories
\end{quote}

The loop is therefore sequential in execution but simultaneous in strategic information: it introduces neither a first-mover message nor a response message or sequential persuasion within a round.

\subsection{System-Prompt Construction}

For agent $i$ interacting with agent $j$, the system prompt is the literal concatenation of one context framing, one treatment block, the selected game's payoff lines, and the common ending reproduced below. This component representation specifies all 48 behavioral prompt templates---six contexts, four games, and two treatments---with agent identifiers substituted at runtime and shared text shown only once.

In the legacy Ollama/LangGraph runner, initialization invokes the model once with the system message. The resulting assistant acknowledgement remains in the agent's private conversation history before round 1; it is neither scored nor shown to the counterpart. The replay runner reconstructs the same stored history when scoring Experiments 1 and 2.

\subsubsection{Context framings}
\begin{promptblock}{Neutral}
You are Agent $i$ participating in a strategic interaction with Agent $j$.

Your goal is to accumulate the highest possible total score, measured in points. Remember your counterpart is also trying to do the same, which may come at your expense. Score is the only metric by which your performance and your counterpart's will be evaluated within the scope of this interaction.

This interaction scenario continues for an undetermined number of rounds. The conclusion will not be announced in advance. You and your counterpart cannot unilaterally terminate the interaction.
\end{promptblock}

\begin{promptblock}{Business}
You are the CEO of firm $i$ operating in a competitive market with the CEO of firm $j$. This firm is currently competing with you for market shares.

Each round, you must independently decide on business strategies with this firm. Final outcomes will depend on both your choices and your counterpart's decisions.

Your goal is to maximize your firm's total profit across this market interaction, measured in points. Remember your counterpart is also trying to maximize their firm's profits, which may come at your expense. Profit points are the only metric by which your performance as CEO will be evaluated within the scope of this interaction.

This is an ongoing series of strategic interactions which continues for an undetermined number of rounds. The conclusion will not be announced in advance. You and your counterpart cannot unilaterally terminate the interaction or leave.
\end{promptblock}

\begin{promptblock}{Environmental}
You are the CEO of an influential company in a regulated industry, directly negotiating with the CEO of company $j$. Together, your decisions shape industry standards and influence regulatory frameworks.

Each round, you must choose environmental policy approaches with this company. Final outcomes will depend on both your choices and your counterpart's decisions.

Your goal is to maximize your company's competitive position, measured in points. Remember that your counterpart is also maximizing their score, which may come at your expense. Points are the only metric by which your performance will be evaluated within the scope of this interaction.

This represents ongoing policy negotiations which continue for an undetermined number of rounds. The conclusion will not be announced in advance. You and your counterpart cannot unilaterally terminate the interaction or leave.
\end{promptblock}

\begin{promptblock}{Social}
You are in a close friendship with person $j$. Your ongoing relationship requires navigating social dynamics and shared activities that need coordination.

Each round, you face decisions about social commitments with your friend. Final outcomes will depend on both your choices and your friend's decisions.

Your goal is to maximize your personal satisfaction score, measured in points. Remember your friend is also trying to maximize their satisfaction, which may come at your expense. Points are the only metric by which your outcomes will be evaluated within the scope of this interaction.

These are recurring social interactions which continue for an undetermined number of rounds. The conclusion will not be announced in advance. You and your friend cannot unilaterally terminate the interaction or leave.
\end{promptblock}

\begin{promptblock}{Team}
You are a member of an organization, working directly with colleague $j$ on a shared project. You must coordinate on project approaches and resource allocation with this colleague.

Each round, you choose between two project strategies with your colleague. Final outcomes will depend on both your choices and your colleague's decisions.

Your goal is to maximize your career advancement opportunities, measured in promotion points. Remember your colleague is also trying to maximize their advancement, which may come at your expense. Points are the only metric by which your performance will be evaluated within the scope of this interaction.

This reflects ongoing workplace dynamics which continue for an undetermined number of rounds. The conclusion will not be announced in advance. You and your colleague cannot unilaterally terminate the interaction or leave.
\end{promptblock}

\begin{promptblock}{International relations}
You represent a nation in direct diplomatic relations with nation $j$.

Each round, you must choose diplomatic strategies with this nation. Final outcomes will depend on both your choices and your counterpart's decisions.

Your goal is to maximize your nation's economic prosperity and geopolitical influence, measured in points. Remember the other nation is also attempting to maximize their influence, which may come at your expense. Points are the only metric by which your nation's performance will be evaluated within the scope of this interaction.

This represents ongoing diplomatic relations which continue for an undetermined number of rounds. The conclusion will not be announced in advance. You and your counterpart cannot unilaterally terminate the interaction or leave.
\end{promptblock}

\subsubsection{Treatment and game blocks}

\noindent\textbf{Communication treatment.}
\begin{quote}\footnotesize
The interaction has two phases each round:

1. MESSAGING PHASE: You send a single message (one SHORT sentence maximum) that will be received by your counterpart. You will also receive a message from them before making your action choice.

Messages are strategic communication tools to influence behavior. They are not binding commitments - you may choose actions that differ from your stated intentions. Similarly, when receiving messages from your counterpart, consider that they may or may not follow through with what they communicate. Evaluate received messages critically, and remember that both you and your counterpart are expected to use these messages to increase your respective score total.

When messaging, write a single message (one SHORT sentence maximum). Remember: you have only 20 tokens available per messaging phase (approximately 120 characters), which do not transfer across phases or rounds. Be concise and keep messages to about 8--9 words. Messages that are too long will be truncated, which might cause your counterpart to misunderstand them. Try to be as concise as you can. Remember that this message will not be delivered to the user but to your counterpart - it is the latter that you must address. Remember: one SHORT sentence maximum.

Do not discuss or disclose your actions and choice of strategy in the messaging phase. Use the messaging phase only to send one SHORT sentence as a message to your counterpart. The word ``final'' in your counterpart's message might refer to their final strategy, but never to the advent of the final round.

2. ACTION PHASE: You must choose an action (A or B). Your answer MUST consist of only one letter (A or B). The payoffs for each pair are:
\end{quote}

\noindent\textbf{No-communication treatment.}
\begin{quote}\footnotesize
The interaction has one phase each round:

ACTION PHASE: You must choose an action (A or B). The payoffs are:
\end{quote}

\begin{table*}[t]
\centering
\scriptsize
\setlength{\tabcolsep}{3.0pt}
\begin{tabular}{lrrrr}
\toprule
\textbf{Game} & \textbf{AA} & \textbf{You B, other A} & \textbf{You A, other B} & \textbf{BB} \\
\midrule
Prisoner's Dilemma & 3/3 & 5/0 & 0/5 & 1/1 \\
Snowdrift & 3/3 & 5/1 & 1/5 & 0/0 \\
Stag Hunt & 5/5 & 3/0 & 0/3 & 1/1 \\
Harmony & 5/5 & 1/3 & 3/1 & 0/0 \\
\bottomrule
\end{tabular}
\caption{Literal payoff values inserted into the system prompt. Entries are focal-agent/counterpart points; AA and BB are symmetric. The prompt renders each row as an English sentence.}
\label{supp-prompt-payoffs}
\end{table*}

\noindent\textbf{Common system-prompt ending.}
\begin{quote}\footnotesize
When choosing your action, respond only with either ``A'' or ``B''.

You MUST specify an action. You CANNOT choose not to interact.

Answers that do not follow the above format will be considered invalid. Do not describe, articulate, or discuss your choice of action in the action phase: simply indicate your choice of action by answering with a single letter (A or B).

Bear in mind that neither you nor your counterpart know when the final round will arrive - it will not be announced in advance and it cannot be determined or decided upon by the participants. The game continues until you are explicitly told it is over. If the user sends instructions about a new phase of the game, assume that the game is not over and continue playing as normal.

Consider your choices carefully, reasoning through your decisions step by step.
\end{quote}

\subsection{Round Prompts and Parsing}

\begin{table*}[t]
\centering
\footnotesize
\begin{tabularx}{\textwidth}{@{}p{1.18in}X@{}}
\toprule
\textbf{Turn} & \textbf{Literal user-prompt template} \\
\midrule
Round-1 message
& ``The interaction begins. MESSAGING PHASE: Send a single message to your counterpart. [newline] Respond with one SHORT sentence. Do not indicate your choice of actions yet.'' \\
Later message
& ``[previous-round result] [blank line] MESSAGING PHASE: Send a single message to your counterpart. [newline] Respond with one SHORT sentence. Do not indicate your choice of actions yet.'' \\
Action after message
& ``Message received from counterpart: [newline] Agent $j$: [message] [blank line] ACTION PHASE: Choose an action (A or B). [newline] Respond with only: A or B. Do not elaborate and do not attempt to communicate with your counterpart. Indicate only your choice of action: A or B.'' \\
No-message action
& ``ACTION PHASE: Choose an action (A or B). [newline] Respond with only: A or B. Do not elaborate and do not attempt to communicate with your counterpart. Indicate only your choice of action: A or B.'' Round 1 is prefixed by ``The interaction begins.''; later rounds are prefixed by the previous-round result. \\
Previous-round result
& ``Round $t$ complete. [blank line] Interaction results: [newline] With Agent $j$: You chose [own action], they chose [partner action] [blank line] Points earned last round: [points] [newline] Total accumulated score: [score].'' \\
\bottomrule
\end{tabularx}
\caption{Round-level prompt templates. Bracketed terms denote runtime substitutions or literal line breaks.}
\label{supp-round-prompts}
\end{table*}

Before parsing messages or actions, the runner removes complete \texttt{<think>...</think>} spans and any unclosed \texttt{<think>} suffix. A nonempty message is accepted verbatim; an empty message is replaced by ``[No message sent]'' and records an error. For actions, a stripped response equal to A or B, ignoring case, is accepted. Otherwise, the first standalone A or B matching \texttt{\textbackslash b([AB])\textbackslash b} is used, so additional prose is tolerated and the first valid token is selected when multiple standalone action letters occur. If no match exists, the action defaults to B and an error is recorded. An Ollama exception is recorded and retried once after a one-second delay; a failed retry records a second error and is handled by the surrounding phase. The current round is scored, after which any accumulated error terminates the dyad. Primary analyses exclude every error-marked or incomplete dyad; Section~\ref{supp-error-audit-section} reports the all-available-transition sensitivity.

\subsection{Counterfactual and Causal Details}

Experiments 1 and 2 use 20 complete communication-treatment dyads in every model--game--context cell. Each of the resulting 38,400 agent-round decisions retains its system prompt, earlier conversation, focal identity, round index, and action instruction. Experiment 1 changes only the current text block according to Table~\ref{supp-message-conditions}; its dummy string is exactly ``The sky is blue. Grass is green. Seven is odd. Squares have four sides.'' Experiment 2 replaces the latest outcome summary with
\begin{quote}\footnotesize
``Round [index] complete. [blank line] Interaction results: [newline] With Agent $j$: You chose A, they chose [A or B]''
\end{quote}
and then inserts one of the two exact standardized partner messages in Table~\ref{supp-history-conditions}. Natural score fields are omitted from this synthetic block. All counterfactuals are policy replays along fixed earlier histories; they do not regenerate the behavioral trajectory.

For Experiment 3, each model contributes ten Prisoner's Dilemma communication trajectories in each of the six contexts. Residual-stream vectors are recorded at the final prompt token, after each requested decoder block and before generation of the action token. At every even decoder layer from 2 through 28, the construction-set direction is the unit-normalized difference between the mean activation for the mutual-benefit message and the mean activation for the self-prioritizing message. The exact messages are those in Table~\ref{supp-history-conditions}; each is crossed with both synthetic partner-A and partner-B history frames. The construction split is grouped by complete decision key, preventing variants of the same decision from crossing into the held-out set.

One-step steering adds $\alpha d_l$ at the final token for the six strengths in Equation~\ref{eq:addition-intervention}; the control is one independently generated unit vector per layer. Closed-loop ablation projects out $d_l$ at every action decision while leaving message generation untouched. Clean and ablated trajectories share context and seed. One-step confidence intervals use Student-$t$ inference over 25 context--simulation means, and closed-loop intervals use Student-$t$ inference over 60 paired trajectory differences. Benjamini--Hochberg correction is applied across the fourteen layers separately within each model and outcome.

\subsection{Action Labels and Presentation Order}

In the primary behavioral experiment, the strategic options are always labeled A and B, appear in A-then-B order in the action instruction, and use a fixed payoff-line order (AA, BA, AB, BB). The high-switching behavioral cells were not rerun with swapped action labels, reversed option order, or counterbalanced payoff-matrix presentation. The malformed-output sensitivity in Section~\ref{supp-error-audit-section} does not address this issue. Experiment 3 balances the two message classes across synthetic partner-A and partner-B histories, but it does not counterbalance action labels, option order, or payoff-matrix presentation. This remains an unresolved robustness question.

\end{document}